\documentclass{article}

\usepackage{arxiv}

\usepackage[utf8]{inputenc} 
\usepackage[T1]{fontenc}    
\usepackage{hyperref}       
\usepackage{url}            
\usepackage{booktabs}       
\usepackage{amsfonts}       
\usepackage{nicefrac}       
\usepackage{microtype}      
\usepackage{lipsum}
\usepackage{bm} 
\usepackage{graphicx}

\usepackage{color}
\newcommand{\rededit}[1]{#1}

\title{A comprehensive spatial-temporal infection model}

\author{
  Harisankar Ramaswamy  \\
  Aerospace and Mechanical Engineering \\
  Viterbi School  of Engineering \\
  University of Southern California \\
  \texttt{hramaswa@usc.edu} \\
   \And
 Assad A. Oberai \\
 Aerospace and Mechanical Engineering \\
  Viterbi School  of Engineering \\
  University of Southern California \\
  \texttt{aoberai@usc.edu} \\
  \And 
   Yannis C. Yortsos \thanks{Corresponding author} \\
  Mork Family Department of Chemical Engineering and Materials Science\\
  Viterbi School  of Engineering \\
  University of Southern California \\
  \texttt{yortsos@usc.edu} \\
}

\begin{document}
\maketitle

\begin{abstract}

Motivated by analogies between the spread of infections and of chemical processes, we develop a model that accounts for infection and transport where  infected populations correspond to chemical species. Areal densities emerge as the key variables, thus capturing the effect of spatial density. We derive expressions for the kinetics of the infection rates, and for the important parameter $R_0$, that include areal density and its spatial distribution. We present results for a batch reactor, the chemical process equivalent of the SIR model, where we examine how the dependence of $R_0$ on process extent, the initial density of infected individuals, and fluctuations in population densities effect the progression of the disease. We then consider spatially distributed systems. Diffusion generates traveling waves that propagate at a constant speed, proportional to the square root of the diffusivity and $R_0$.
\rededit{Preliminary analysis shows a similar behavior on the effect of stochastic advection.}

\end{abstract}

\keywords{Infection Models \and Chemical Processes \and Wave Propagation \and COVID-19}

\section{Introduction}

\rededit{Understanding the spread of infectious diseases,} where infection is from human to human, is of significant interest, greatly relevant to epidemics, such as the one the world is experiencing today with COVID-19. A plethora of epidemiologic models have been proposed, developed and tested, {\rededit{e.g. see  \cite{kermack1927contribution,croccolo2020spreading,harko2014exact} among many others}}. These are based on three essential components: (i) Identifying different populations (typically, susceptible, infected, and recovered or perished), and their extensions to demographic or health history subcategories; (ii) Describing ways by which the various populations come in proximity with one another; and (iii) Postulating rates by which members from one population \rededit{convert } into another. 

Widely used is the celebrated SIR model \cite{kermack1927contribution,anderson1979population}, which captures essential aspects of contagion using three populations: susceptible (denoted by $S$), infected (denoted by $I$) and recovered (including perished) (denoted by $R$). The original model, and many of its recent variants, account for items (i) and (iii) above, but not for (ii). More importantly, the SIR representation is in terms of the total populations, rather than their areal densities (people/area), which are expected  to be the most important contagion variables. Such models also ignore spatial transfer \rededit{limitations, as they assume infinitely large diffusion or mobility}.  

Another category \rededit{of models} is based on computing individual trajectories of typical members of the various populations and on subsequently postulating probabilities of collisions and infection \rededit{e.g. \cite{burghardt2020unequal} and many others}. These agent-based models implicitly account for spatial density effects through various mobility, propagation and collision rules. In particular, for specific individuals, whose trajectories and potential contacts are known, one can predict or retrace past infection paths. \rededit {Importantly, } this forms the basis of contact tracing methods. 

A different and, we believe, more fundamental way to model the problem, is to follow a chemical reaction engineering approach, in which one can cast contagion spreading in terms of corresponding transport and reaction models. This \rededit{is a continuum approach that } relies on the following analogies: populations map into chemical species; densities (specifically, areal densities) into molecular concentrations; infection rates into chemical reaction rates (where mass action kinetics apply); and spatial transport into advective and diffusive (or dispersive) fluxes. Then, one can express the relevant population (species) balances using partial differential equations description. It is the objective of this paper to create such a methodology to model the spread of an epidemic in general, with application to COVID-19, in particular. Our work parallels related previous continuum models, \rededit{such as} \cite{noble1974geographic} \rededit{or particularly the very recent work of \cite{viguerie2020diffusion},} but differs in a number of aspects. \rededit{For example, the recent work of \cite{viguerie2020diffusion} uses a compartmental modeling approach using continuum mechanics, that allows one to obtain results including diffusion and reaction similar to the work here. An important difference of the present work is that the view of contagion in the form of chemical reaction processes allows non-trivial additional insights,  including the ab-initio representation of the kinetics of the infection, as well as the effect of mobility through advection.}

\rededit{As in any continuum model, the} fundamental underlying assumption is that one can homogenize population distributions by defining continuum variables, \rededit{expressed as continuous functions of position and time}. This  allows postulating continuum conservation laws in terms of their rate of change, transport and reaction. We recognize limitations in the analogy between transport and interaction of human populations on the one hand and molecular transport and reaction on the other. \rededit{Indeed, given} the relatively sparse areal densities of human populations, compared to molecular densities, the law of large numbers to obtain  well-defined  \rededit{spatial} averages (homogenization) \cite{ross2009first,bensoussan2011asymptotic}, \rededit{may in fact not be in effect}. \rededit{In such cases}, the equivalent distributions might be more akin to rarefied gas dynamics \cite{sharipov2012ab}. \rededit{Further,} human movement is influenced by behavioral drivers, rather than random walks, and it is likely that, e.g. a diffusion or dispersion process for human population mobility would require a description different than Brownian motion. These alternatives will be pointed out where appropriate \rededit{but will not be further pursued. Indeed, we believe that the analogies made  here are useful, novel and instructive, and allow} important new insights on fundamental spatial aspects of the problem.

The commonly used SIR model arises from the present formulation in the limit of "perfect mixing", namely the absence of spatial gradients, over specific control areas (e.g. a building, a manufacturing plant, a school, a city, a state, a country, etc.) \rededit{and assume perfect mixing, induced by fast diffusion or advection}. \rededit{The chemical engineering analogue is the} “batch reactor” model \cite{fogler2010essentials}. We will examine its validity, \rededit{as} infection rates, depending crucially on spatial (areal) density, lead to spatial non-uniformities, while \rededit{ finite transport} leads to infection waves. Both impact the stationary assumptions in the SIR models. First, we find that the key variable $R_0$ \rededit{does not actually remain} constant during the process, contrary to the commonly \rededit{made} assumption. Instead, it is found to depend on the areal spatial density and to decrease as a function of the extent of the contagion.  We \rededit{then} consider the solution of spatially-dependent problems by including diffusion in both 1-D (rectilinear) and  in 2-D \rededit{(including radial symmetry)} geometries. We discuss the asymptotic emergence of traveling infection waves, the speed and shape of which are found to be independent of  geometry, \rededit{but to also depend on the relative importance of transport, with an intensity that is lower, hence corresponding to a smaller value of $R_0$, when diffusion is weaker}. \rededit{Advection is discussed in the context of its most important effect, namely in a heterogeneous, stochastic environment. As in related problems \cite{sreenivasan2019turbulent}, it leads to an effective macrodispersion, which typically dominates over diffusion and leads to enhanced mixing of populations, hence an increase in the intensity of the contagion.} 

The paper is organized as follows: We first proceed with constructing \rededit{ the analogous}  chemical reaction and transport process. After recasting all conservation equations in dimensionless form, we derive the associated key parameter $R_0$. Then, we consider the solution of a number of specific problems, \rededit{which range} from a “batch reactor” model (the SIR equivalent), to the propagation of infection in \rededit{both} time and space (including both 1-D and 2-D domains). Infection waves are found to depend only on $R_0$, or its effective equivalent, and to be largely independent of initial conditions. This lack of influence of initial conditions on the shape of the infection curve (in space and/or in time) is notable, as it signals universal scaling properties, a property already implicitly assumed in many SIR-type models.

\section{Formulation}

Consider the conservation of mass of the three \rededit{areal} quantities of interest, \rededit{namely,} susceptible, infected, and recovered (including perished), and associate to them three equivalent “chemical species”. \rededit{In writing the corresponding mass balances, we associate species mobility through both  advection and diffusion mechanisms, and infection and recovery rates through equivalent chemical reactions. Then, the equations read as follows} 
\begin{eqnarray}
\frac{\partial \rho_i}{\partial T} + \nabla \cdot (\bm{q_i} \rho_i) =  - \nabla \cdot ( \bm{D}_i) + r_ i, \qquad i =S,I, R. \label{eq:condensity}
\end{eqnarray}
where $\rho_i$  is the population density (number/area) \rededit{for species $i$}, $T$ is dimensional time, \rededit{spatial coordinates are in dimensional notation}, the spatial coordinates are only in two dimensions, $\bm{q}_i$ (if any) is the advective velocity vector \rededit{of species $i$}, $\bm{D}_i$ is the diffusive (or dispersive) flux, and $r_i$  is the net reaction rate of species $i$, that convert populations into one another due to infection and recovery. \rededit{The overall species balance dictates  $r_S+r_I+r_R=0$.}

\rededit{In (\ref{eq:condensity}) the area is defined as the surface over which populations reside, work or interact (e.g. work floor area, geographical urban area, etc.} The advective velocity $\bm{q}_i$ denotes \rededit{an advective} transport mobility term. While we allow in (\ref{eq:condensity}) the three velocities to be different, in the remainder we will only take $\bm{q}_i= \bm{q}$ for all populations. This common velocity is assumed independent of $\rho_i$ or its gradients, although it can be a function of space and/or time. Next, define
\begin{eqnarray}
\rho_S+\rho_I+\rho_R = \rho,
\end{eqnarray}   
and      
\begin{eqnarray}
\bm{D}_S+ \bm{D}_I+ \bm{D}_R= \bm{D} 
\end{eqnarray}                                                                                       
where $\rho$ is the total population density (total number/area) and $\bm{D}$ is the total diffusive (or dispersive flux), to obtain
\begin{eqnarray}
\frac{\partial \rho}{\partial T} + \nabla \cdot (\bm{q} \rho) =  - \nabla \cdot ( \bm{D})  \label{eq:con_tot_density}
\end{eqnarray}
as the equation governing the evolution of total density.

\subsection{Diffusion}
How to represent the diffusive (or dispersive) fluxes requires some discussion. For a typical Fick’s law type of diffusion \cite{bird1961transport}, one may take
\begin{eqnarray}
\bm{D}_i= - D \nabla \rho_i,     \label{eq:difffick}               
\end{eqnarray}
where we defined a constant diffusion (or dispersion) coefficient $D$, and assumed that all species are indistinguishable as far as their physical properties is concerned. \rededit{Then, Equation (\ref{eq:difffick}) yields $\bm{D}= - D \nabla \rho$ and (\ref{eq:con_tot_density}) becomes}  
\begin{eqnarray}
\frac{\partial \rho}{\partial T} + \nabla \cdot (\bm{q} \rho) =  \nabla \cdot ( D \nabla \rho)  \label{eq:con_tot_fick}
\end{eqnarray}
This suggests that the total population density $\rho$, in addition to being advected, also diffuses in the direction of a negative spatial gradient. While possible and perhaps even likely in human dynamics, the notion that the overall density diffuses is contrary to the common continuity equation for \rededit{physical phenomena in} fluids, e.g. incompressible fluids \cite{lamb1993hydrodynamics}. \rededit{This contradiction is removed if one takes  the  alternative approach and defines} instead 
\begin{eqnarray}
\bm{D}_i= - D \rho \nabla (\rho_i/\rho),         \label{eq:difflat}           
\end{eqnarray}
\rededit{Then,} the more familiar form emerges 
\begin{eqnarray}
\frac{\partial \rho}{\partial T} + \nabla \cdot (\bm{q} \rho) =  0 \label{eq:con_tot_lat}
\end{eqnarray}
Diffusive fluxes of the type (\ref{eq:difflat}) do in fact arise in random walks on non-uniform lattices, in the limit when the lattice becomes continuous \cite{coifman2006diffusion,hoffmann2019spectral}. While either (\ref{eq:difffick}-\ref{eq:con_tot_fick}) or (\ref{eq:difflat}-\ref{eq:con_tot_lat}) can be taken to describe diffusive transport, in the examples \rededit{to be discussed }below we will only assume the more familiar continuity equation versions (\ref{eq:difflat}) and (\ref{eq:con_tot_lat}). This has the additional advantage that in the absence of advection, the total density is time-independent and a stationary function of space, which is useful for the solution of many problems of interest, since the equations now become decoupled. \rededit{It is important to note that while the overall density in (\ref{eq:con_tot_lat}) does not diffuse, the individual species (populations) continue to do so via (\ref{eq:difflat}).}

\rededit{As remarked above diffusion} or dispersion can occur by different than a Brownian motion type random walk, e.g. by Levy flights (where the distribution of diffusion steps has a heavy tail, thus allowing for occasional, although rare, large steps) \cite{mandelbrot1982fractal}. Modeling such motions can be handled by an integro-differential equation description through fractional derivatives  \cite{chaves1998fractional,del2003front}. While worth considering, such an approach will not be pursued here. \rededit{We will also not explore additional, possibly interesting, cases, e.g. of non-linear diffusion, in which the diffusion coefficients are functions of the individual densities, e.g. see \cite{viguerie2020diffusion}. In crowded environments of high density, diffusion will certainly be affected by the total density, $\rho$, although it will likely be independent of the \textit{individual} density fractions. In problems of constant overall density, therefore, those types of non-linear effects will be absent. Conversely, one must consider cases where the advection velocity is a stochastic variable. The resulting fluctuations will be expected to lead to enhanced macrodispersion, e.g. as in turbulence or in stochastic transport in porous media \cite{sreenivasan2019turbulent,gelhar1983three}, dominating over diffusion, thus contributing to enhanced spreading of the contagion. }

Diffusion coefficients can be estimated by considering the ratio of the square of the average radius of a random walk over an associated time interval. For instance, for an office type environment where  random walks may have a mean radius of $10 m$, over an $8-hr$ period, one obtains $ D \approx 10^{-2}  m^2/s$. \rededit{By comparison, molecular diffusion coefficients in gases are about three orders of magnitude smaller}. Dispersion at much larger scales can be obtained by inference  from the spatial spread of the epidemic. As we describe later, diffusion leads to infection waves, whose propagation speed depends on the assumed diffusion coefficient. We show that values of the order of $ D \approx 10^3 m^2/s$ or higher are derived in order to match the observed propagation velocities. Clearly, such large values reflect \rededit{not only diffusion, but rather }an aggregate mix of transport activities (including \rededit{macrodispersion} through advection).   

\subsection{Reaction Rates}

Consider, next, contagion rates. Following our chemical reaction analogy, we postulate the following two chemical reactions 
\begin{eqnarray}
S+ I &\to& 2 I   \label{eq:reac1} \\
I &\to& R       \label{eq:reac2}                                                      
\end{eqnarray}
that convert the three species to one-another. The stoichiometric coefficient 2 in the RHS of (\ref{eq:reac1}) indicates that  one member of infected species $I$ is produced as a result of its interaction with one member of susceptible species $S$. In turn, species $I$, consumed in reaction (\ref{eq:reac2}), leads to species $R$,  produced in one-to-one stoichiometry. Both reactions are irreversible.

Applying mass action kinetics \cite{erdi1989mathematical} in the reactions (\ref{eq:reac1}) and (\ref{eq:reac2}) provides expressions for the reaction rates in terms of the respective concentrations or densities \rededit{namely,} 
\begin{eqnarray}
r_S=-K \rho_S \rho_I,\qquad  r_I=K \rho_S \rho_I-\Lambda \rho_I, \qquad r_R= \Lambda \rho_I,   \label{eq:rates}
\end{eqnarray}
where we introduced the reaction rate constants $K$ and $\Lambda$. Equations (\ref{eq:rates}) are the same as those for the SIR model, except that here the rates are correctly expressed in terms of areal densities, rather than in terms of the total populations, \rededit{as} inaccurately assumed in typical models. \rededit{This has significant implications on the kinetic parameters, as discussed below. The two kinetic constants have different dimensions, $\Lambda$ expressed in inverse time, and $K$  in inverse (time $\times$ number/(length)$^2$).}  

We hasten to note that in reality a more fine-grained model would be applicable, the reaction rates also depending on demographics and/or health conditions of the individual species. Such fine-graining is possible \rededit{with the present framework,} by further partitioning the populations into subgroups \cite{liu2019dynamical}. The corresponding description would then be cast in terms of an equivalent ``multicomponent mixture'' \cite{wesselingh2000mass}, where equations (\ref{eq:condensity}) are expressed in terms of an extended species vector, with reactions (\ref{eq:reac1})-(\ref{eq:reac2}) extended appropriately, possibly involving a product of a reaction matrix with the species vector. A related multicomponent description, although based on a different continuum model using applied mechanics methodologies, was presented in \cite{viguerie2020diffusion}. For simplicity, this generalization will not be further considered here. 

Relevant to linear (“first-order”) reactions, $\Lambda$ has dimensions of inverse time, a typical value being $1/14 \; days^{-1}$ \cite{noble1974geographic, gog2014spatial, viguerie2020simulating}. This kinetic parameter expresses the rate at which infected individuals recover (or die), on average, and it is intrinsic to the infected fraction. This is not the case for non-linear (e.g. “second-order”) reactions, like those involved in the rate of generation of new infections (reaction (\ref{eq:reac1})). As reflected in the kinetic parameter $K$, infection kinetics will depend on the duration, method and type of human-to-human contact, the protective gear (PPE) of susceptible and infected individuals, various biological and physiological variables, the ambient environmental conditions (room air conditioning), spatial distancing and other parameters. Most controllable among these factors are the frequency and degree of of encounters (collisions) between individuals, as well as the intensity of interaction.  

Possible approaches to estimating $K$ include kinetic theory models (e.g. similar to Maxwell-Boltzmann models), where the kinetic parameter is inversely proportional to the molecular mean free path (average length between collisions), itself \rededit{being }inversely proportional to density \cite{hirschfelder1964molecular}. At least in some domain \rededit{of the values of $\rho$}, $K$ should be increasing with spatial density, \rededit{a feature} ignored in previous SIR models. Directly applying Maxwell-Boltzmann-type kinetic theories is \rededit{unrealistic } in the present context \rededit{however,} and must be \rededit{further} refined: Human encounters are not elastic collisions, and typically last over finite time intervals. More importantly, effects of spatial distancing, a recognized key to the kinetics of contagion, must be \rededit{accurately} captured. Indeed, it is by now widely accepted that infection rates are negligible for densities below a limiting value (corresponding to a separation of $2 m$ or $6 ft$, as also supported by fluid mechanical models of droplet propagation,  and recommended in health policy guidelines  \cite{wells1934air,bourouiba2014violent,xie2007far}). 

\rededit{We incorporate all these aspects by postulating} the following dependence
\begin{eqnarray}
K(\rho) = \left\{ \begin{array}{cc} 0, & \rho < \rho_0  \\ K_0 F(\frac{\rho - \rho_0 }{\rho_1 - \rho_0}), & \rho \ge \rho_0 \end{array} \right. \label{eq:defk1} 
\end{eqnarray}
where $F(x)$ is an increasing function of $x$, $F(0)=0$, $F(1)\approx 1$. \rededit{Here, the threshold } value $\rho_0 \approx 0.1 m^{-2}$ represents the density below which the reaction rate is negligible,  while the upper limit $\rho_1 \approx 1 m^{-2}$ corresponds to a maximum “packing” density. \rededit{In } (\ref{eq:defk1}) we separated spatial distancing effects, included through $\rho$, from factors, such as biological, environmental, facial covering, etc., which enter through $K_0$ only (e.g. with $K_0$ decreasing substantially as facial covering is applied).  

Equation (\ref{eq:defk1}) captures both spatial distancing and biological and environmental effects. Upon more careful inspection\rededit{, however} the density dependence must be further modified to not include the recovered fraction component, since that fraction does not affect contagion (assuming infection immunity for all recovered individuals). Therefore, the way density enters in (\ref{eq:defk1}) \rededit{must be modified }by replacing it with $\rho(1-r)$ (and where we defined  $r= \frac{\rho_R}{\rho}$, see also below). The new expression (\rededit{namely, }replacing $K(\rho)$ with $K(\rho (1-r))$) \rededit{, then} reads
\begin{eqnarray}
K= K(\rho, r) = \left\{ \begin{array}{cc} 0, & \rho (1-r) < \rho_0  \\ K_0 F(\frac{(1-r)\rho - \rho_0 }{\rho_1 - \rho_0}), & \rho(1-r) \ge \rho_0 \end{array} \right. \label{eq:defknew} 
\end{eqnarray}
When the rate of infection is relatively low (and $r \ll 1$), the \rededit{above } correction is negligible. For strong infection rates, on the other hand, it can be quite significant, as \rededit{also shown} later in the paper.

Some additional remarks are pertinent: The above assume that an infected individual can infect a susceptible one at the same constant rate. This is  true either for asymptomatic, infected individuals, or for those who do not exhibit symptoms until a few days following infection. It is not true, however, when infected individuals are isolated. Nonetheless,  the previous formalism \rededit{can still be applicable}: \rededit{For example, }assume that the infected species $\rho_I$ is further subdivided into one category containing asymptomatic individuals (denoted by $A$, with corresponding density $\rho_A$) and another containing quarantined individuals (denoted by $Q$, with corresponding density $\rho_Q$). The associated infection reaction rate \rededit{now reads} $K \rho_S \rho_A$. Based, further, on the reasonable assumption that the percentage of those infected, but are asymptomatic or with mild symptoms, is a fixed fraction (e.g. $a$) of the total fraction of the infected, the infection rate becomes $K a \rho_S \rho_I$. This is of the same dependence as before, hence the previous holds, \rededit{ subject to modifying  the kinetic constant with the} parameter $a$. The same reasoning (with an additional parameter included) holds when infected individuals are contagious, but not identified as such, until sometime after infection. On the other hand, our approach \rededit{does not} as easily account for correlations between infected and susceptible individuals, for example when susceptible individuals have increased contact, hence higher probability of infection, with specific infected individuals related to them, e.g. by family, work or other proximity relations.     

A final remark relates to the practice of reporting area-wide averages (e.g. for states or countries). Given that almost all areas will never on average reach the minimum density required for infection (e.g. $0.1 m^{-2}$), area-wide averages over substantially heterogeneous density distributions are not very informative. Rather, distinguishing high-density areas (e.g. urban places, stadiums, schools, retirement homes, etc.) \rededit{or events attracting high densities}, from low-density ones (e.g. farms, rural) is essential. \rededit{Such } reporting of more fine-grained area statistics is much more informative. Connecting the transmission of infection between areas of different density is \rededit{readily feasible with} the present formalism, which includes spatial transport and/or diffusion, and where $K$ is space-dependent. \rededit{These features are}  further discussed below. 

\subsection{Dimensionless formulation} 

We \rededit{will next proceed with rendering equations (\ref{eq:condensity}) dimensionless. First, we introduce the notation $s=\frac{\rho_s}{\rho}$, $i=\frac{\rho_i}{\rho}$, and $r=\frac{\rho_r}{\rho}$, namely we normalize species densities by $\rho$. Then, we use equation (\ref{eq:difflat}) for the diffusion terms and equations (\ref{eq:rates}) for the reaction rates. Finally, we dimensionalize time by $1/\Lambda$, space by a characteristic external length scale $L$, $K$ by $K_0$, and velocities by a characteristic velocity $U$, and use lower case symbols for all dimensionless variables. We obtain the following set of differential equations}
\begin{eqnarray}
\frac{\partial s}{\partial t} + Da \bm{v}\cdot\nabla s  - C \nabla (\ln \rho ) \cdot \nabla s &=&   \nabla \cdot ( C \nabla s) -R_0 (\rho, r) s i  \label{eq:s1} \\
\frac{\partial i}{\partial t} + Da \bm{v}\cdot \nabla i  - C \nabla (\ln \rho ) \cdot \nabla i &=&   \nabla \cdot ( C \nabla s) + R_0 (\rho, r) s i   - i \label{eq:i1} \\
\frac{\partial r}{\partial t} + Da \bm{v}\cdot\nabla r  - C \nabla (\ln \rho ) \cdot \nabla r &=&   \nabla \cdot ( C \nabla r) + i \label{eq:r1} \\
\frac{\partial \rho}{\partial t} + Da \nabla \cdot (\bm{v} \rho)  &=&   0 \label{eq:density_nondim}
\end{eqnarray}
\rededit{Here}, we defined the dimensionless Damkohler number $Da=\frac{U}{L\Lambda}$, the dimensionless diffusion number $C=\frac{D}{\Lambda L^2 } = \phi^{-2}$, and the rescaled velocity $\bm{v}$. \rededit{We recognize} $\phi$ as the Thiele modulus of the chemical reaction engineering literature  \cite{fogler2010essentials}. From this dimensionless formulation arises \rededit{naturally} the most important parameter $R_0$, commonly associated with the \rededit{spreading of epidemics}, namely  
333\begin{eqnarray}
R_0(\rho,r) = \frac{K_0 \rho }{\Lambda} \kappa(\rho, r) \label{eq:defrho1} 
\end{eqnarray}
where 
\begin{eqnarray}
\kappa(\rho,r) = \left\{ \begin{array}{cc} 0, & (1-r)\rho < \rho_0  \\ F(\frac{(1-r)\rho - \rho_0 }{\rho_1 - \rho_0}), & (1-r)\rho \ge \rho_0 \end{array} \right. \label{eq:defkap1} 
\end{eqnarray}
Equation (\ref{eq:defrho1}) shows that \rededit{even in spatially homogeneous systems,} $R_0(\rho, r)$ is not constant, as typically assumed, but also depends both on the areal density and on \rededit{a measure of the intrinsic} extent of the process $r$. For example, the ratio of the final value  $R_0(\rho,r_{\infty})$ to its initial $R_0(\rho, 0)$ can be approximated  by  
\begin{eqnarray}
\frac{R_0(\rho,r_{\infty})}{R_0(\rho, 0)}=\frac{(1-r_{\infty})\rho-\rho_0}{\rho-\rho_0}\approx 1-r_{\infty}  
\label{eq:ratio} 
\end{eqnarray}
assuming $\rho_0/\rho \ll 1$ \rededit{and a linear function for $F$}. For consistency in the remainder, we will \rededit{denote} the  $R_0$ dependence of various solutions to the problem, e.g. of $r_{\infty}$,  through its value at the onset of the process, \rededit{namely, through} $R_0(\rho, 0)$. 

The finding that $R_0(\rho, r)$ is not constant is consistent with spatial distancing health policy guidelines. To our knowledge this property has not been noted before, even though it has important implications on the prediction of infection results, as shown below.      

For completeness, we also remark that had we used the different version for diffusion (\ref{eq:difffick}), instead of (\ref{eq:difflat}), equations (\ref{eq:s1}), (\ref{eq:i1}) and (\ref{eq:r1}) would still remain in effect, subject to multiplying by a factor of $2$ the third term on the LHS of each equation. 

\section{Applications}

\rededit{We are now in a position to consider a number of interesting applications. With the exception of the analytical results for the batch reactor problem, all other } numerical results were obtained by solving Equations (\ref{eq:s1})-(\ref{eq:r1}) using a finite-elements method \rededit{with the} weak form of these equations implemented and solved in the Fenics software package \cite{alnaes2015fenics}. The spatial domain was discretized using linear finite elements, and the resulting equations were integrated in time using the trapezoidal rule. The mesh size and the time-step were chosen to \rededit{be} sufficiently small such that further refinement did not lead to significant changes in the resulting solution.

\subsection{Perfect Mixing: The SIR Model}
\rededit{Consider first } the zero-dimensional (“batch reactor”) problem, with no input or output \rededit{in the control volume, at } conditions of perfect mixing. \rededit{Dependent variables, as well as density, are not functions of the spatial coordinates, the implication being that mobility effects, such as advection and diffusion are sufficiently large to homogenize the system. The partial differential equations (\ref{eq:s1})-(\ref{eq:r1}) then become ordinary differential equations in time. Using $(\dot{})$ to denote time derivatives, we obtain the following system of ordinary differential equations}
\begin{eqnarray}
\dot{s} (t) &=& -R_0 (\rho,r) s i \label{eq:s2} \\
\dot{i} (t) &=&  R_0 (\rho,r) s i - i \label{eq:i2} \\
\dot{r} (t) &=&  i \label{eq:r2}
\end{eqnarray}
subject to the closure 
\begin{eqnarray}
s + i + r &=& 1  \label{eq:clos1} 
\end{eqnarray}
and the initial conditions
\begin{eqnarray}
i(0)=i_0, \qquad s(0) \equiv s_0=1-i_0, \qquad  r(0)= 0. \label{eq:ic1} 
\end{eqnarray}
\rededit{We must point out  that even  though the overall density  in such an SIR-like model is time-independent, spatial effects do enter through the effect of density (hence of the extent of contagion) on $R_0$ (equations (\ref{eq:s2}), (\ref{eq:i2})). }. 

Equations (\ref{eq:s2})-(\ref{eq:r2}) produce non-trivial results when an initial, even infinitesimally small, seed of infected individuals ($i_0$) is present. An analytical solution is possible. Substitute (\ref{eq:r2}) into (\ref{eq:s2}) and integrate to give:  
\begin{eqnarray}
s  = s_0 \exp(- \int_0^r R_0(\rho,r') dr' ) \label{eq:s}
\end{eqnarray}
thus, 
\begin{eqnarray}
i  = 1 - r - s_0 \exp(- \int_0^r R_0(\rho,r') dr' ) \label{eq:i}
\end{eqnarray}
Further substitution into (\ref{eq:r2}) gives 
\begin{eqnarray}
\dot{r} (t)  = 1 - r - s_0 \exp(- \int_0^r R_0(\rho,r') dr' ) \label{eq:temp}
\end{eqnarray}
which can be integrated to \rededit{provide }the final solution
\begin{eqnarray}
t = \int_0^r \frac{du}{1 - u - s_0 \exp(- \int_0^u R_0(\rho,r') dr') }. \label{eq:fin} 
\end{eqnarray}
\rededit{Results are shown in the sections to follow. In (\ref{eq:clos1}) we have tacitly assumed that at the onset of the process, all individuals are susceptible. This assumption precludes the possibility of immune individuals, e.g. via vaccination. We will discuss this important point later below.} We also remark, in passing, that by expanding the exponential in (\ref{eq:temp}) in a Taylor series, assuming constant $R_0$ and keeping the first three terms in the expansion, leads to a Riccati equation
\begin{eqnarray}
\dot{r} (t)  = i_0- r(1 - s_0 R_0(\rho,0)) -\frac{s_0 R_0(\rho,0)^2}{2} r^2
\end{eqnarray}
accurate for small $R_0$. \rededit{This equation can be used as a simpler model for quick insights. Riccati equations have been used as a simpler  alternative to the SIR description \cite{marmarelis2020predictive,fokas2020covid,fokas2020two}}.

\subsubsection{Infection Curves, Herd Immunity and Effective $R_0$}

\begin{figure}[h!]
    \centering
    \includegraphics[width=0.6\textwidth]{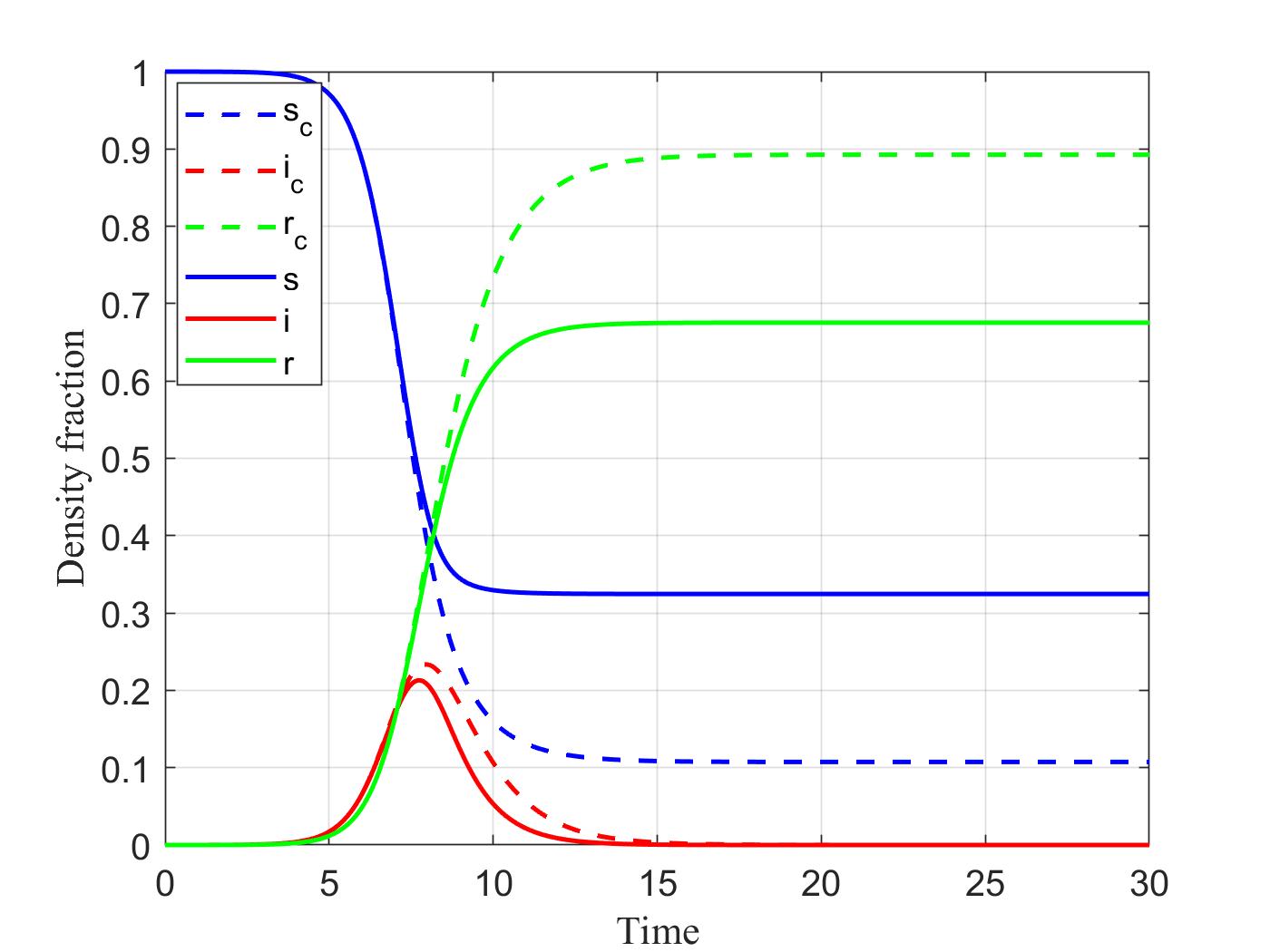}
    \caption{Variation of susceptible ($s$), infected ($i$) and recovered ($r$) fractions as a function of non-dimensional time for the case when $R_0$ varies following equation (\ref{eq:defkap1}) (solid lines) with $R_0 (\rho ,0)= 2.5 $, and for the case of a constant value of $R_0=R_0 (\rho ,0)= 2.5 $  (dashed lines), which is the SIR assumption. Note the substantial difference between the results obtained between a constant and a variable $R_0$. The initial infected fraction is $i(0) = 10^{-5}$.}
    \label{fig01}
\end{figure}

\begin{figure}[h!]
    \centering
    \includegraphics[width=0.6\textwidth]{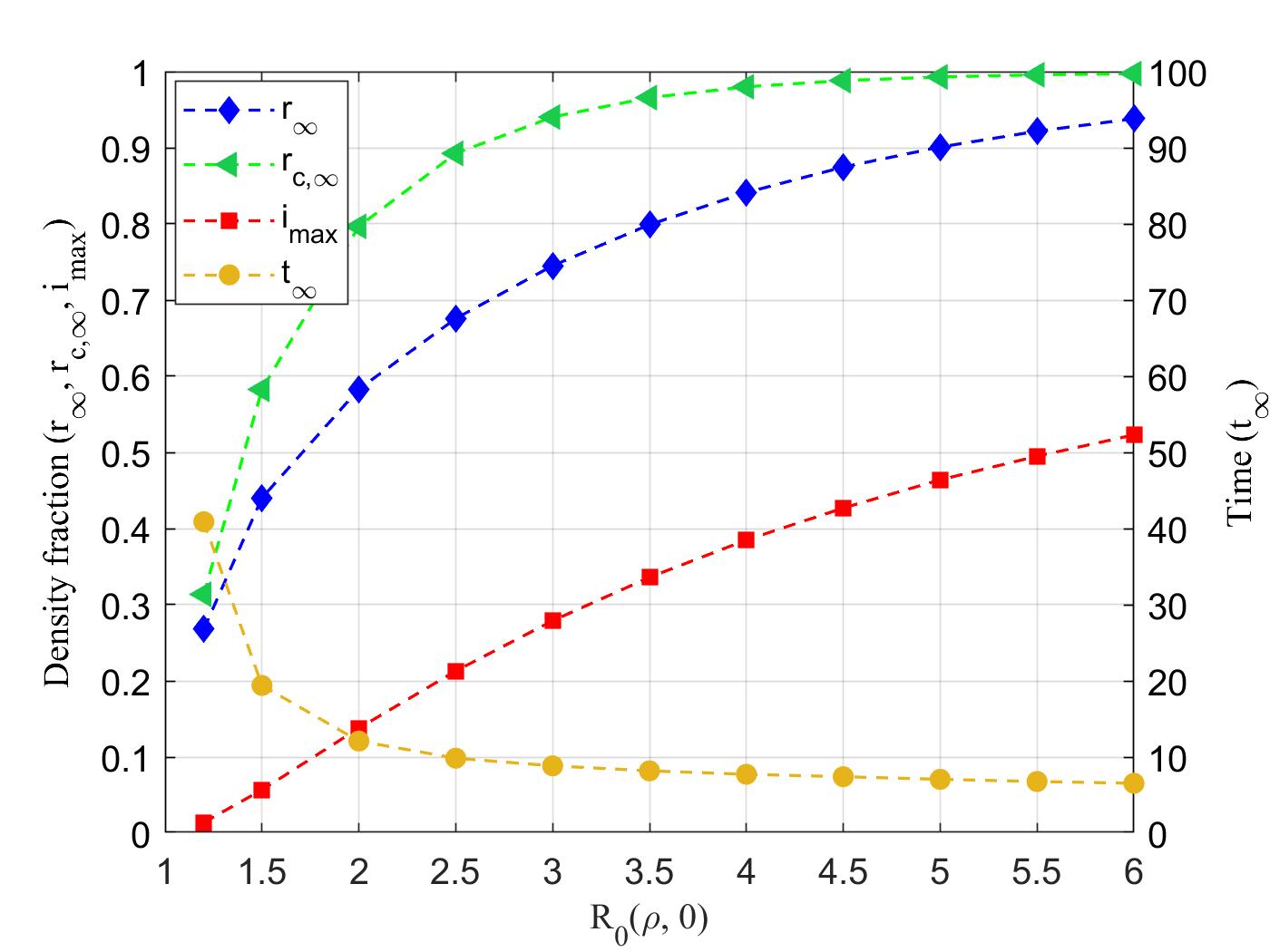}
    \caption{Variation of herd immunity $h$ (denoted here as $r_\infty$), maximum infection fraction ($i_{max}$) and the duration of a pandemic ($t_\infty$) as a function of $R_0(\rho,0)$. Plotted also is the herd immunity ($r_{c,\infty}$) assuming a constant $R_0$. The initial infection fraction is $i(0)=0.001$.} 
    \label{fig02}
\end{figure}

The solutions of \rededit{equations (\ref{eq:s})-(\ref{eq:fin})}  are shown in Figure \ref{fig01}. We first remark that from Equation (\ref{eq:i2}), $R_0 (\rho,0)=1$ is the boundary demarcating two regions, where an initial infection either decays ($R_0 (\rho,0)<1$) or grows ($R_0 (\rho,0)>1$). We will focus on the latter case ($R_0 (\rho,0)>1$). Plotted in Figure \ref{fig01} is the time variation of the three different populations for $R_0 (\rho ,0)= 2.5$, as well as of the curves obtained under the SIR assumption of constant $R_0$ (taken throughout the process to be equal to $R_0 (\rho ,0)= 2.5$). The  infection curves are of a similar shape, but \rededit{with significantly larger values if a constant $R_0$ value is used.} \rededit{One could define an} effective constant $R_0$, using the average 
\begin{eqnarray}
R_0\equiv \frac{\int_0^{r_\infty} R_0(\rho,r') dr'}{r_\infty}
\label{eq:rhoaver}
\end{eqnarray}
\rededit{Its relation to} $R_0 (\rho,0)$ is inferred from Figure \ref{fig02}, which for example shows that to a constant value of $R_0=2.5$ throughout the process corresponds a twice as large \rededit{initial} value of $R_0 (\rho,0)=4.5$.

\rededit{Parameter $R_0$ is typically interpreted as the number of new infections caused on average by an infected individual. Such an interpretation is ill-defined, however, in that it does not specify the interval of time, over which such infection will occur. The latter depends on the frequency and the ultimate exposure by susceptible populations to the infected individual. If we assume \rededit{for simplicity }an exponential rise of infections with constant $s$ and constant $R_0$ in (\ref{eq:i2}), the interpretation of $R_0$ as the number of new infections caused on average by an infected individual requires an exposure time of $t_{\rm exp}=\frac{lnR_0}{R_0-1}$, e.g. $t_{\rm exp}= 0.549$ for $R_0=3$, or  $t_{\rm exp}= 0.46$ for $R_0=4$, both roughly corresponding to one week, for typical values. This has significant implications for {\it{super-spreader}} events in that for an infected individual to infect a number of others, during the short period of time of the event, the corresponding prevailing value of $R_0$ must be substantially larger (of the order of tens) than 1.}      

A related quantity is the length of the epidemic, here approximated by  
\begin{eqnarray}
t_\infty = \int_0^{r_e} \frac{du}{1 - u - s_0 \exp(- \int_0^u R_0(\rho,r') dr') }.  
\end{eqnarray}
where $r_e=0.99 r_\infty$, and  $r_\infty$ is the solution of (\ref{eq:rinfty}). Results are shown in Figure \ref{fig02}. We note that the dependence of $t_\infty$ on $R_0 (\rho,0)$ is monotonically decreasing, with smaller values of $R_0 (\rho,0)$ (and lower infection levels) resulting into a longer duration, as long as 30, in dimensionless time,  corresponding roughly to 15 months or longer. A more typical duration, but with higher infection rates, is about 12 dimensionless time units (6 months). The fact that the epidemic lasts \rededit{longer} at smaller values of $R_0$ is counter-intuitive, and calls for the need to educate the public that epidemics on the border of being under control will last longer \rededit{ than intuitively expected.} 

Plotted in Figure \ref{fig02} is also the maximum in the infected fraction, $i_{\rm max}$, which is an increasing function of $R_0 (\rho,0)$, as expected. \rededit{Equation (\ref{eq:i2}) shows that the maximum is reached when $R_0 (\rho,r^*) s^*=1$, where superscript * corresponds to the values at that point. }   

Of importance is the concept of {\it herd immunity}, denoted here by $h = r_\infty$, and defined as the asymptotic value of the recovered individuals, $r_\infty$, \rededit{based on a given value of $R_0(\rho,0)$}. It is the solution of the equation
\begin{eqnarray}
1 - h - s_0 \exp(- \int_0^{h} R_0(\rho,r') dr' ) = 0. \label{eq:rinfty}
\end{eqnarray}
Results are shown in Figure \ref{fig02}. We note that $h$ is an increasing function of $R_0 (\rho,0)$ (with $h \rightarrow i_0$ as $R_0 (\rho,0) \rightarrow 0$). Also plotted in the same figure is the herd immunity calculated under the SIR assumption of a constant $R_0 = R_0 (\rho,0)$. The corresponding values are significantly higher, even for relatively mild rates of infection. \rededit{It is important to realize that herd immunity has the commonly accepted interpretation, if and only if the corresponding value of $R_0$ that resulted in the arresting of the contagion is maintained at long times. Otherwise, and if behavioral changes cause $R_0$ to increase, following an erroneous understanding of herd immunity, then contagion will commence again (e.g. second or third waves, etc). We can explain this further in terms of the stability of the asymptotic state.}

\begin{figure}[h!]
    \centering
    \includegraphics[width=0.6\textwidth]{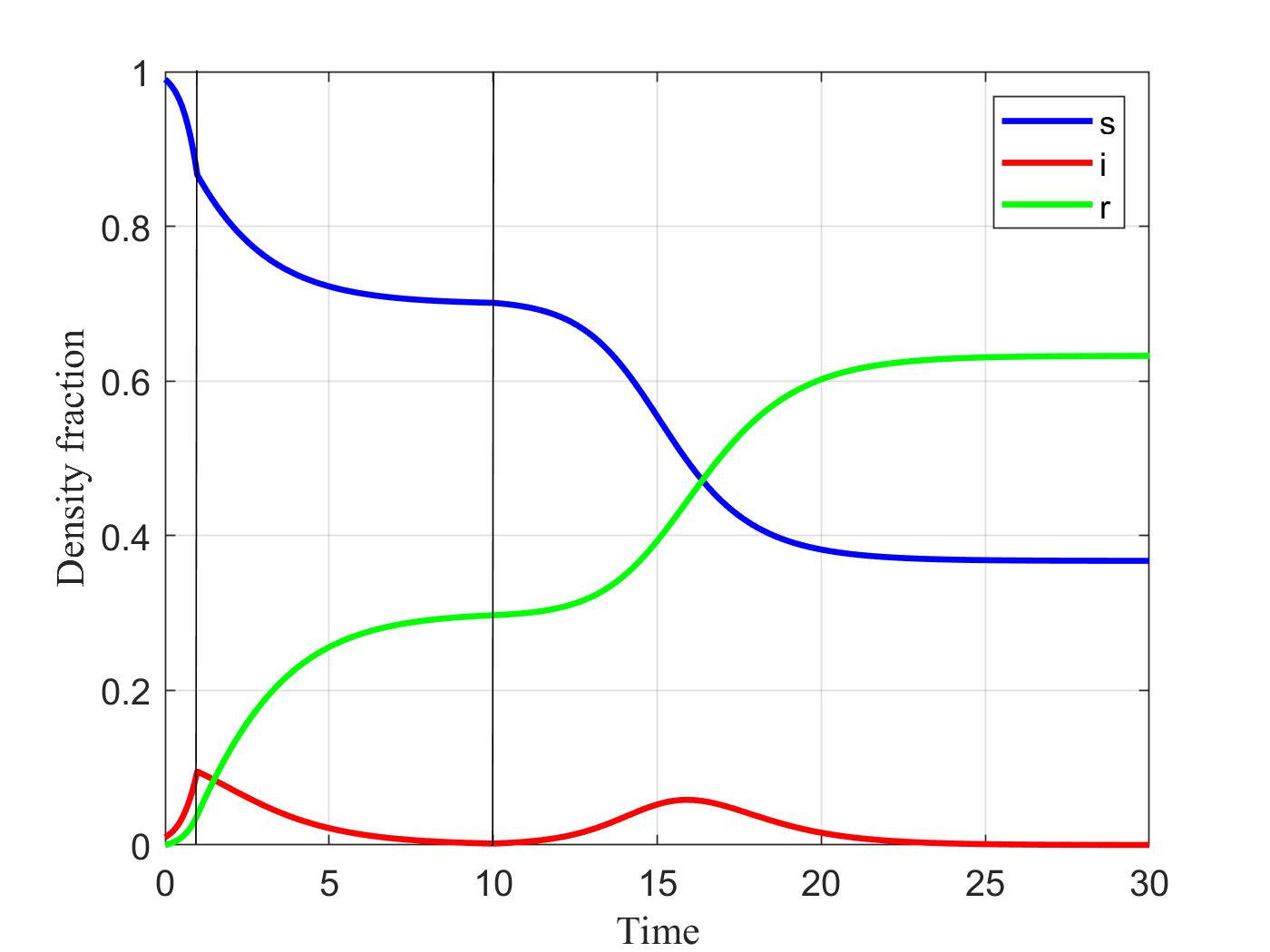}
    \caption{Variation of susceptible ($s$), infected ($i$) and recovered ($r$) fractions as a function of non-dimensional time with $R_0(\rho,0) = 3, t \in (0,1), R_0(\rho,0) = 0.8, t \in (1,10), R_0(\rho,0) = 3, t \in (10,30) $. The initial infected fraction is $i(0) = 0.01$.}
    \label{fig03}
\end{figure}

Consider a small resurgence of infections after an asymptotic state (\rededit{and its associated herd immunity} is reached. As long as the asymptotic condition $R_0 (\rho,h)(1-h )<1$ is satisfied (which happens to always be the case), such fluctuations will decay  exponentially fast, as shown in equation (\ref{eq:i2}): the final state is asymptotically stable. This will not be the case, however, when the fluctuation is instead in $R_0 (\rho, h)$, e.g.  when a new value, e.g. $R_0'$,   such that $R_0' (1-h )>1$, sets in. For example, this could be the result of relaxations on spatial distancing, and/or of abandoning due caution, after the asymptotic state is reached. \rededit{Under such conditions}, there will be an eruption of new cases, resulting into a new spreading of infections (a “second wave”), which will in turn reach a new asymptotic state, and a correspondingly new, higher herd immunity. Figure \ref{fig03} demonstrates such a second wave. The figure shows three different regimes: An initial one with $R_0 (\rho,0)=3$ for which infection grows; a second regime following the imposition of restrictions at $t=1$ (which in the specific example leads to $R_0=0.8$), and which results in the "flattening of the curve" with a corresponding herd immunity of $0.3$; and a regime in which the relaxation of restrictions at $t=10$ and a return to a higher value of $R_0 (\rho,0)=3$, leads to a second wave. The new epidemic lasts at least as long as the first one, and contributes to substantial new infections (almost double the initial). This lack of structural stability of the asymptotic state derives from the fact that the infection reaction (\ref{eq:reac1}) is auto-catalytic, and leads, at least initially, to exponential rises. It illustrates the importance of closely adhering to a consistent, and as low as possible, value of $R_0$, for desired herd immunity levels to be sustained.

\rededit{We close this part by noting that condition \rededit{$R_0 (\rho,r_\infty)(1-r_\infty )<1$, also written as} $R_0 (1-h)<1$, namely $h>1-\frac{1}{R_0}$ delineates the state of "herd immunity" typically understood by the public. In the absence of a vaccine or of other means (e.g. spatial distancing) of keeping $R_0$ small, herd immunity means that a non-trivial fraction of the population under consideration must be infected, and recovered (or perished). For example, for a behavior corresponding to $R_0=4$, the associated herd immunity is 0.75, and likewise for other values. More generally, assume that a fraction of the initial population has been immunized, e.g. through vaccination. A way to model this problem would be to assign this fraction initially to the "recovered" population fraction, namely to take instead of (\ref{eq:ic1}) the following initial conditions}

\begin{eqnarray}
i(0)=i_0, \qquad s(0) \equiv s_0=1-i_0- r_0, \qquad  r(0)= r_0. \label{eq:icVacc} 
\end{eqnarray}
\rededit{Then, the previous solution applies, by using   in all integrals over $r$, in equations (\ref{eq:i})-(\ref{eq:fin}), the lower limit of  $r_0$ instead of  $0$. In particular, we deduce that for the infection to not spread, and thus be contained from its onset the following must apply: $R_0 (\rho,r_0)s_0<1$, which in view of (\ref{eq:icVacc}), means $R_0 (\rho,r_0)(1-r_0)<1$. This is the same condition as above for the onset of herd immunity, namely $h> 1-\frac{1}{R_0}$. For example, if pre-pandemic behavior in terms of spatial distance and facial covering has an associated value of $R_0=5$, then the corresponding requirement on immunization is $80>\%$.}

\subsubsection{Universal Scaling}
\begin{figure}[h!]
    \centering
    \includegraphics[width=0.6\textwidth]{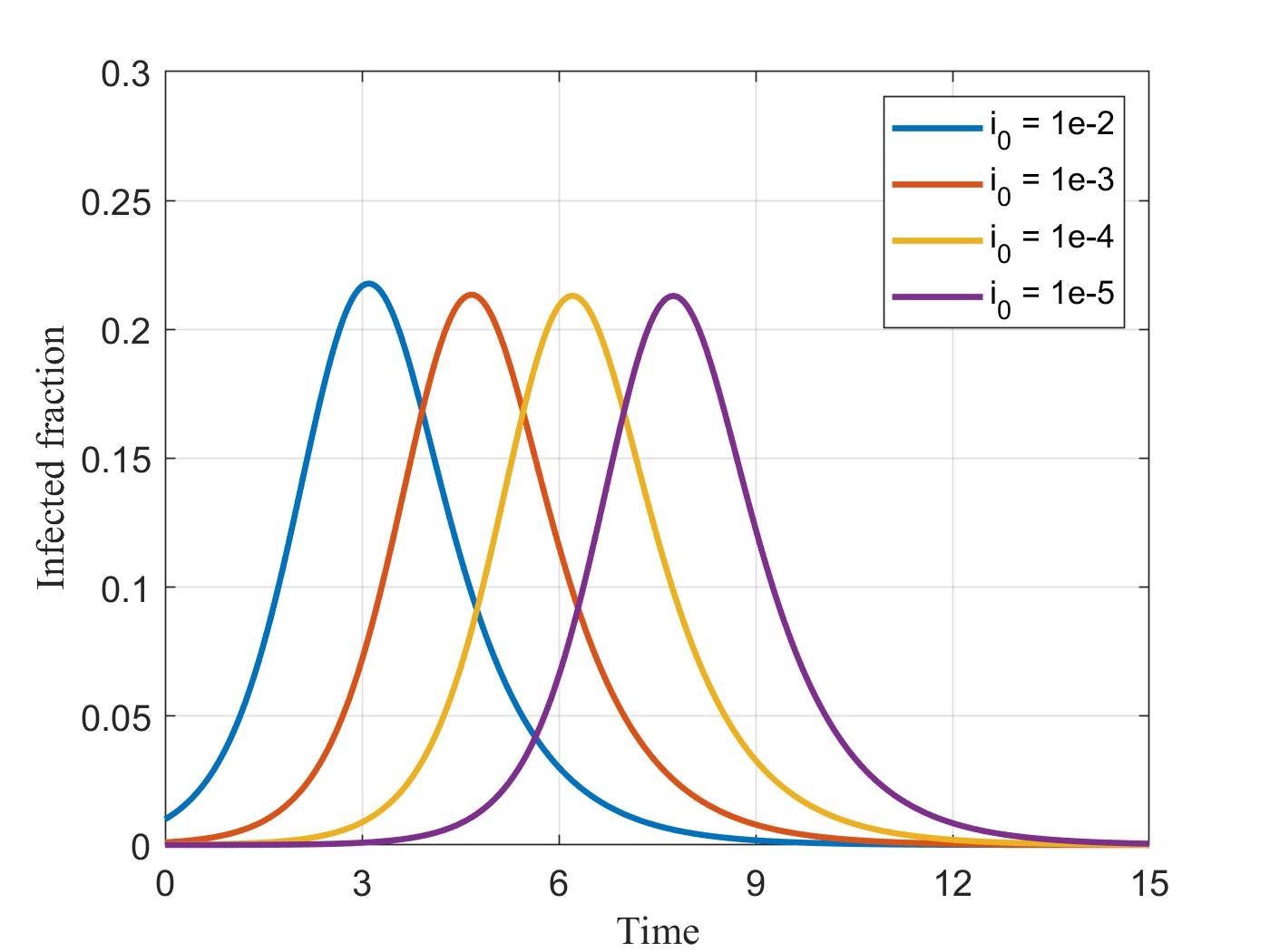}
    \caption{Variation of infected ($i$) fraction as a function of non-dimensional time for different values of the initial infected fraction $i_0$ and $R_0(\rho,0) = 2.5$.}
    \label{fig04}
\end{figure}

The auto-catalytic nature of reaction (\ref{eq:reac1}) raises the additional question as to whether or not the infection curves depend on parameters other than $R_0$. Consider, first, the effect of the initial condition $i_0$. Figure \ref{fig04} is a plot of the infection curve for different values of the initial condition and for $R_0 (\rho,0)= 2.5$. For the typical values considered,  a decrease in the initial condition fraction leads to a shift in the infection curve $i(t)$ to the right, but otherwise produces shapes that are almost identical. The shift is approximately 2 non-dimensional time units for each decrease in the initial condition by a factor of 10. 

\rededit{We can  explain these findings by considering} the early-time solution of equation (\ref{eq:i2}). At small times we obtain  
\begin{eqnarray}
i(t) \approx i_o \exp ((R_0 (\rho,0) - 1) t) = \exp ((R_0 (\rho,0) - 1) (t - t_0)), \label{eq:earlytime} 
\end{eqnarray}
where  $t_0 = - \frac{\log (i_0) \ln (10)}{R_0 (\rho,0) - 1} \approx - 2.3 \frac{\log (i_0)}{R_0 (\rho,0) - 1} $. This  confirms the existence of a time shift $t_0$, and of the same magnitude as in the figure, consistent with the simulations. For the same reasons, the maximum infection fraction $i_{max}$ is independent of the initial condition $i_0$, assuming that the latter remains relatively small. The invariance in the shape of the infection curve, and the fact that a decrease in the number of initial infections only acts to delay the onset of infection, are significant from a health policy perspective: containing the initial number of infections provides a non-trivial interval of time to contain the infection, e.g. by educating the public to modify behavior, thus to lower $R_0$ and ultimately mitigate the intensity of the incoming infection. Absent such preparation or behavior modification, will negate any beneficial effect of the lower number of initial infections: A corresponding contagion wave will emerge, ultimately dependent only on $R_0$.    

\subsubsection{Imported Infection}

The same conclusions apply for “imported infection”. With this term, we refer to the case where a control area (e.g. a country), originally without any infected individuals, receives a constant influx of infected and susceptible individuals over a finite (and small, reflecting a prompt public authority response) time interval $\tau$, following which such influx is stopped (e.g. via a “flight ban”).  By integrating equation (\ref{eq:i1}) over spatial dimensions and assuming perfect mixing we obtain to leading order
\begin{eqnarray}
\dot{i}(t) = (R_0 (\rho,0) -1) t +j_i
\label{import}
\end{eqnarray}
valid for $0< t <\tau$. Here, we denoted the influx rate by $j_i$ and also assumed that the corresponding changes in density are negligible, as long as $\tau$ is small.  Solving equation (\ref{import}) subject to the initial condition $i(0)=0$, we obtain, to leading-order (small $\tau$) 
\begin{eqnarray}
i=j_i t + \cdots
\end{eqnarray}
\rededit{essentially providing an equivalent initial condition, $i(0)=i_0=j_i \tau$. It follows that as long as $R_0 (\rho,0) \tau \ll j_i$, the impact of “imported infections” is simply to nucleate the process, and the problem reverts to the previous. As before, the net impact of a ban is to provide additional time (e.g. a month) for the necessary public health measures to be implemented and to lead to as small a value of $R_0 (\rho,0)$ as possible, thereby limiting or even preventing an inevitable spread of infection.} In the absence of additional such prophylactic measures, a ban can only serve to delay the onset of contagion.  

\subsection{Fluctuations}

The batch (SIR-like) model rests on the assumption that all profiles are spatially uniform. We explore this assumption in the presence of fluctuations or other non-uniformities, by considering the solution of a typical 1-D problem \rededit{in the absence of advection, in which case we have} 
\begin{eqnarray}
\frac{\partial s}{\partial t}  - C \frac{\partial \ln \rho }{\partial x} \frac{\partial s}{\partial x} &=&   C \frac{\partial^2 s}{\partial x^2} -R_0 (\rho, r) s i  \label{eq:s3} \\
\frac{\partial i}{\partial t}  - C \frac{\partial \ln \rho }{\partial x} \frac{\partial i}{\partial x} &=&   C \frac{\partial^2 i}{\partial x^2} + R_0 (\rho, r) s i   - i \label{eq:i3} \\
\frac{\partial \rho}{\partial t} &=&   0 \label{eq:density_nondim3}
\end{eqnarray}
subject to no-flux conditions at the two ends, $\frac{\partial s}{\partial x} = \frac{\partial i}{\partial x} = \frac{\partial \rho}{\partial x} = 0$, at $x = \{0,1\}$. Consider the case where the initial fluctuations are variable, but the density is uniform,
\begin{eqnarray}
i(x,0) = i_0 (1+ \epsilon g(x)), \qquad s(x,0) = 1-i(x,0), \qquad \rho(x,t) = \rho_m, \label{eq:init_fluc}
\end{eqnarray}
or where the non-uniformity is in the density profile only
\begin{eqnarray}
i(x,0) = i_0 , \qquad s(x,0) = 1-i(x,0), \qquad \rho(x,t) = \rho_m (1+ \epsilon g(x)),
\end{eqnarray}
where $\epsilon \ll 1$. Expecting that diffusion ($C>0$) will help to smooth spatial non-uniformities, we will focus in this section on the solution in its absence ($C=0$). Thus, any effects to arise will correspond to  the spatial averaging  of an ensemble of batch problems.  

\subsubsection{Stability}
When $C=0$ the relevant equations revert to (\ref{eq:s2})-(\ref{eq:r2}). Consider a small perturbation expansion and denote means by superscript bar and fluctuations by superscript prime to obtain
\begin{eqnarray}
\frac{\partial \bar{s}}{\partial t}  &=&    -R_0 (\rho_m) (\bar{s} \bar{i} + \overline{s'i'})  \label{eq:save} \\
\frac{\partial \bar{i}}{\partial t}   &=&  + R_0 (\rho_m) (\bar{s} \bar{i} + \overline{s'i'})   - \bar{i} \label{eq:iave} 
\end{eqnarray}
where we ignored the effect of the fluctuations on $R_0$. By further taking the representation $s' = \epsilon i_0 \sigma(t) g(x)$ and $s' = \epsilon i_0 \eta(t) g (x)$, the fluctuations to leading order satisfy
\begin{eqnarray}
\frac{d \sigma}{d t}  &=&    -R_0 (\rho_m) (\bar{s} \eta  + \bar{i} \sigma )  \label{eq:sprime} \\
\frac{d \eta }{d t}   &=&  + R_0 (\rho_m) (\bar{s} \eta  + \bar{i} \sigma )   - \eta \label{eq:iprime} 
\end{eqnarray}
with corresponding initial conditions  $\sigma(0) = -1$ and $\eta(0) = 1$.

We first note that fluctuations contribute to the rate expression, Equations (\ref{eq:save})-(\ref{eq:iave}), through the mean of the product  $ \overline{s'i'} $. This contribution is negative and proportional to the square of the amplitude of the fluctuations  $ \overline{(i_0 \epsilon g(x))^2}$. A first  conclusion is that  fluctuations lower the effective rate of the infection reaction (although only to order  $\epsilon^2$ assuming that the fluctuations remain bounded). Adding diffusion will further reduce any such impact.  

\begin{figure}[h!]
    \centering
    \includegraphics[width=0.6\textwidth]{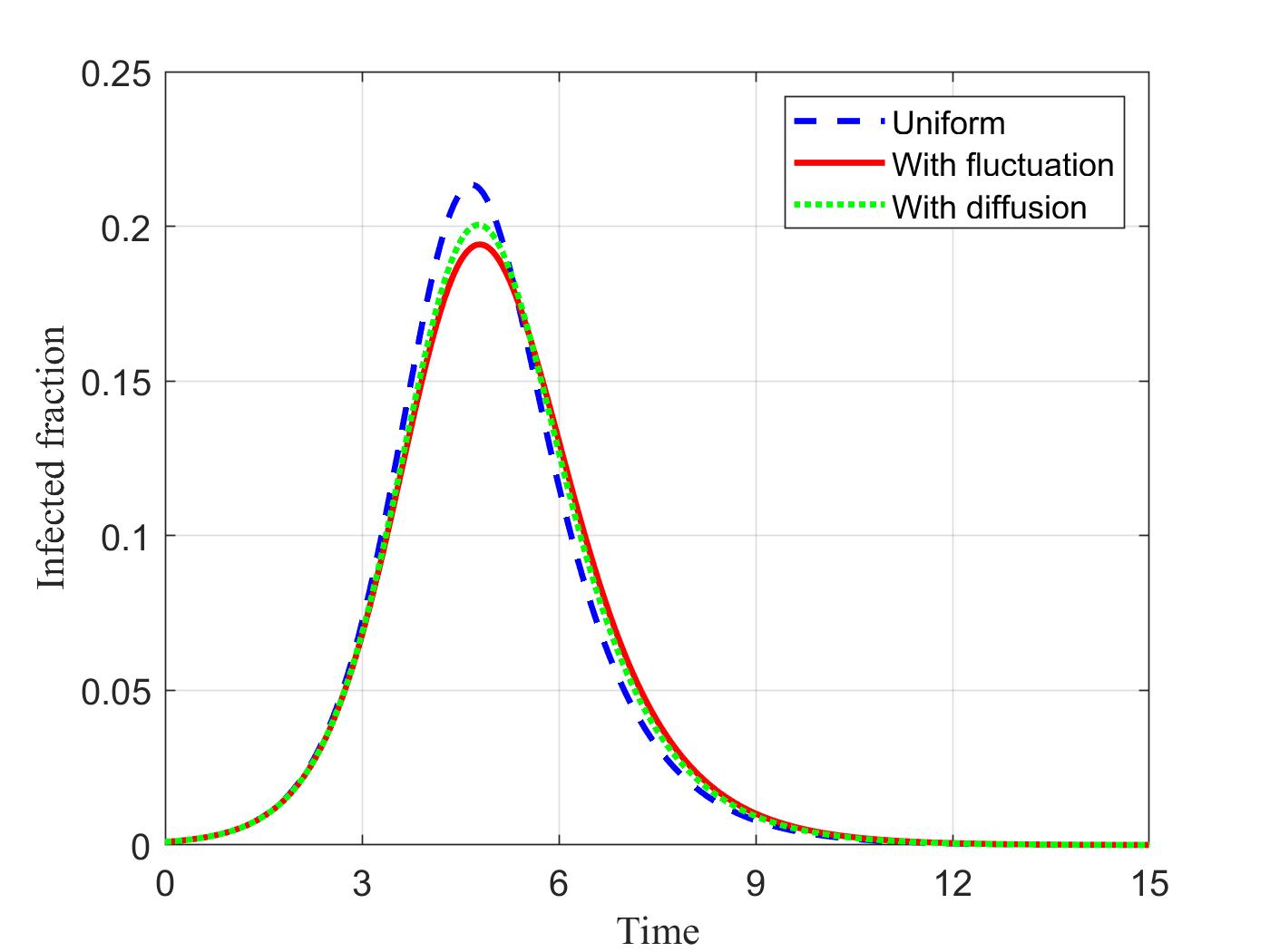}
    \caption{Variation of the ensemble average of the infected ($i$) fraction as a function of non-dimensional time for $R_0(\rho,0) = 2.5$. The initial condition is equation (\ref{eq:init_fluc}) with $i_0 = 0.05$, $g(x) = \sin (x)$, and $\epsilon = 0.04$.}
    \label{fig05}
\end{figure}

To determine whether or not fluctuations are bounded, we must find the eigenvalues $\omega$ of the matrix of the linear system (\ref{eq:sprime})-(\ref{eq:iprime}),
\begin{eqnarray}
\omega^2 + (R_0 (\rho_m) \bar{i} - R_0 (\rho_m) \bar{s} +1) + R_0 (\rho_m) \bar{i} = 0. 
\end{eqnarray}
This equation has two negative eigenvalues as long as  
\begin{eqnarray}
R_0 (\rho_m) \bar{i} - R_0 (\rho_m) \bar{s} +1 >0. \label{eq:eigcon}
\end{eqnarray}
Equations (\ref{eq:s2})-(\ref{eq:r2}) show that for some time before the infection fraction $i$ reaches its maximum, and always after that time, condition (\ref{eq:eigcon}) is indeed valid. One concludes that fluctuations will remain bounded, if not altogether decay, hence they have little effect on the average behavior. This result is demonstrated in Figure \ref{fig05}, where we plot ensemble average responses. The curves obtained are practically the same either in the presence or in the absence of fluctuations.  

\subsubsection{Spatial Heterogeneity}
The results are quite different, however, when spatial heterogeneity is stronger. Here, the interpretation of the composite average must be done with a careful understanding of the underlying heterogeneities.

\begin{figure}[h!]
    \centering
    \includegraphics[width=0.6\textwidth]{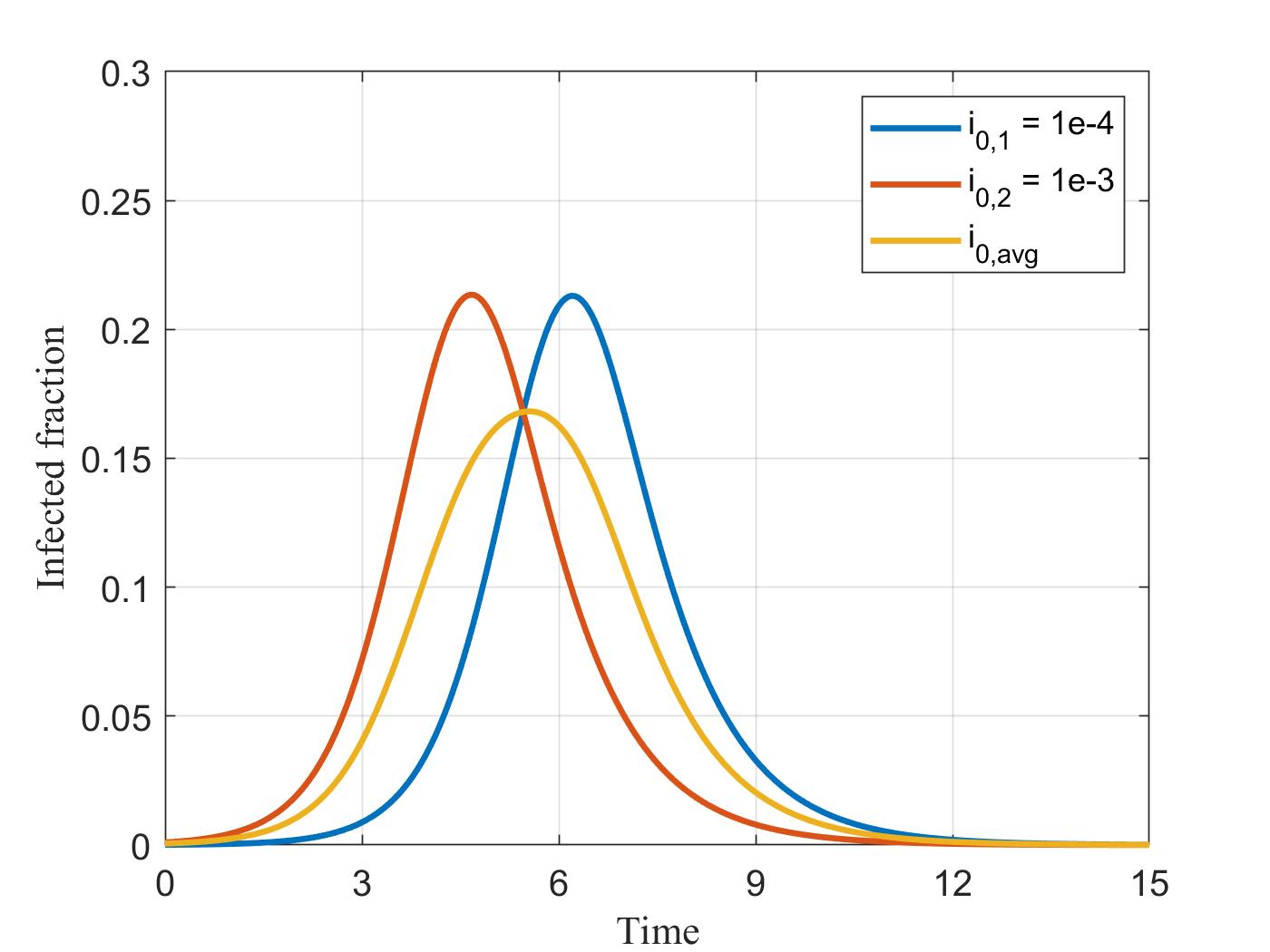}
    \caption{The infected ($i$) fraction curves as functions of non-dimensional time for the two different regions with different initial infections, and the composite curve ($R_0(\rho,0) = 2.5$).}
    \label{fig06}
\end{figure}

Consider, first, two regions with the same density, but with different initial conditions that differ by an order of magnitude (e.g. $i_1=10^{-4}$ and $i_2=10^{-3}$)
\begin{eqnarray}
i(x,0) = i_1 H(\xi-x) +  i_2 H(x  - \xi)
\end{eqnarray}  
where $0 < \xi <1$ and $H$ is the Heaviside step function. From the previous analysis we expect that the region with higher initial infections will respond faster, the other response trailing by a time lag (e.g. \rededit{as suggested in } Figure \ref{fig04} and Equation (\ref{eq:earlytime})). Accordingly, the composite behavior will be controlled initially by the region with the larger number of initial infections, but at later times by the second region. Figure \ref{fig06} shows that this is indeed the case. 

More interesting, perhaps, is the case where the area of interest consists again of two regions, e.g. "urban" in the interval $0<x<\xi$ and "suburban" in the interval $\xi <x<1$. We are interested in the infection curve under the conditions of a “commute” between the two regions: During the day, and for a certain time interval of duration $\lambda$ (expressed as a fraction of the 24-hr period), the density in the urban region, $\rho_u$, is high, and in the suburban region it is negligible; while during the remaining part of the day and at night, the density in the urban region is zero, and in the suburban is $\rho_s$. These two densities are related by the conservation equation $\rho_s =\rho_u \frac{\xi}{1-\xi}$. This setting is intended to model the commute between two regions (“home” and “work”, or “home” and “school”) where we expect significantly different $R_0$ values. \rededit{For further simplification, we will take}  $\xi \ll 1$, and simply assume $R_0=0$ for conditions at “home”.   

\begin{figure}[h!]
    \centering
    \includegraphics[width=0.6\textwidth]{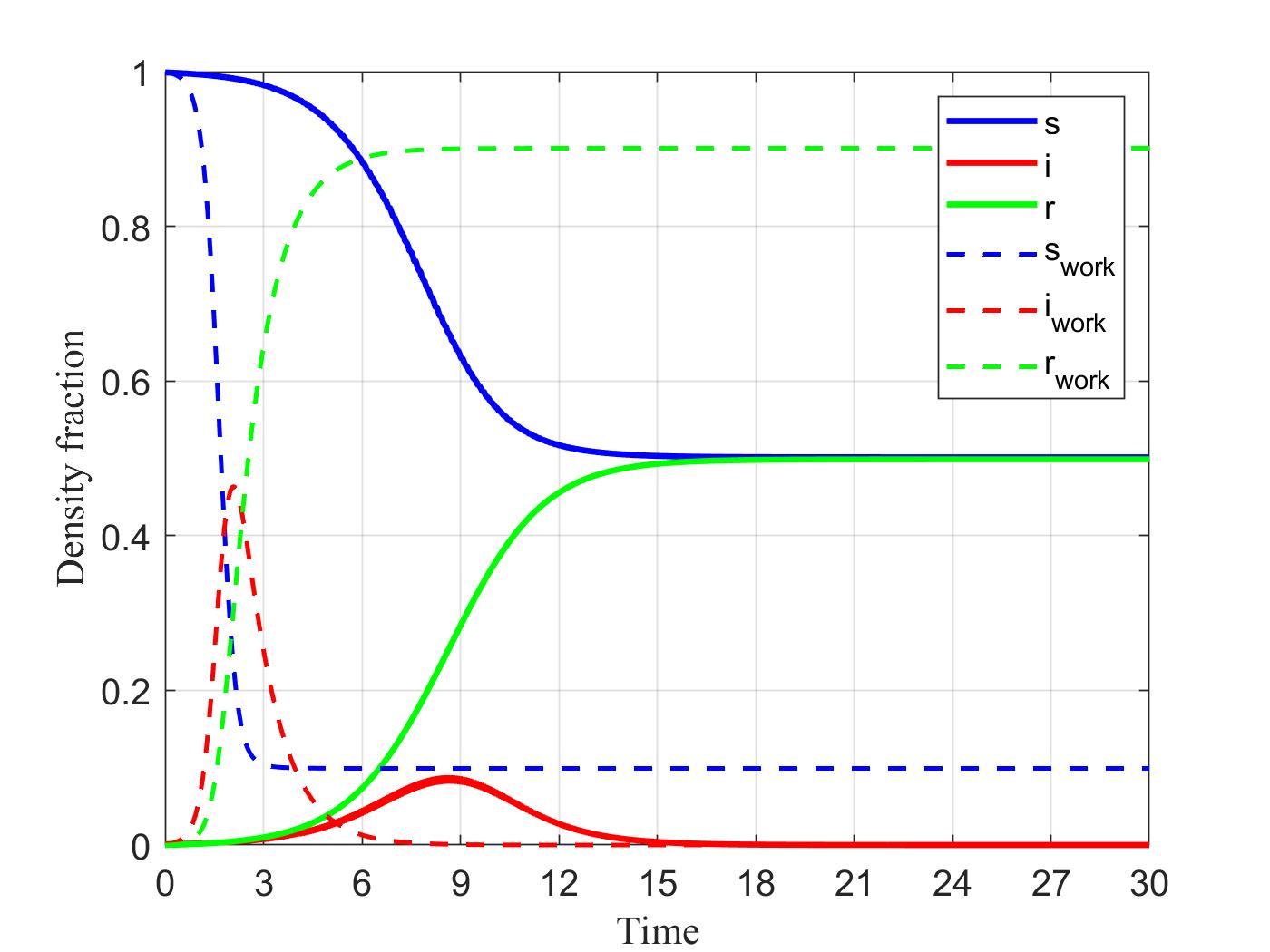}
    \caption{The infection curves for the case of a commute between two regions of high and low values of $R_0$ ($R_{0,w}(\rho, 0) = 5$) and  exposure frequency of $\lambda=1/3$ (solid lines). Plotted also are the infection curves corresponding to a uniform $R_{0,w}(\rho, 0) = 5$ (equivalently to $\lambda=1$) (dashed lines). The initial infection fraction is $i(0)=0.001$.} 
    \label{fig07}
\end{figure}

This problem can be solved based on the detailed transport and reaction equations derived earlier, that include advection and diffusion. A simpler alternative is to represent it  as two "batch reactors", whose $R_0$ values oscillate between the two values, $R_{0,w} (\rho, r)$ and $0$, when the population is at “work” (or “school”) or at “home”, respectively. The results of this hypothetical “commute”, for an 8-hr work period ($\lambda = 1/3$), are shown in Figure \ref{fig07}. Superimposed are also the results that would have emerged if the "work" parameter $R_{0,w} (\rho, r)$ was applied in both regions (or, equivalently, \rededit{when} $\lambda=1$). 
As expected for this example, while commuting leads to a much lower effective value $R_{0,eff} (\rho, 0)=1.67$, it does not suppress infection:  As long as  $R_{0,eff} (\rho, 0)>1$ infection will occur, although by reducing exposure (decreasing $\lambda$) the \rededit{resulting effective rates are significantly lower}. Indeed, we can show \rededit{using the asymptotic method of ``two-timing'' \cite{holmes2012introduction} that for all values of $\lambda$ the corresponding effective value is the arithmetic mean, $R_{0,eff} (\rho, 0)= \lambda R_{0,w}$.} This suggests that there is a critical exposure value $\lambda_{crit}= \frac{1}{R_{0,w}}$, below which contagion is suppressed. These  findings, are new and must be interpreted with due caution, \rededit{as they are} subject to the assumptions made.          

\subsection{The Effect of Transport}
The preceding sections dealt with applications in zero-dimensional space (batch reactors or their ensembles), \rededit{namely under the assumption of fast diffusion or other transport }. Transport (via advection, diffusion or dispersion) plays two roles: One is to reduce the effect of small spatial non-uniformities, as discussed above. The other is in the direction of further spreading the infection spatially. This section will consider \rededit{such mobility effects, starting first with diffusion}.   

\subsection{Effect of Diffusion }
\rededit{In the absence of advection ($Da \bm{v}  = \bm{0}$) and for a constant diffusion coefficient,} the relevant equations become 
\begin{eqnarray}
\frac{\partial s}{\partial t}   - C \nabla (\ln \rho ) \cdot \nabla s &=&   C  \nabla^2 s  -R_0 (\rho, r) s i  \label{eq:s4} \\
\frac{\partial i}{\partial t}  - C \nabla (\ln \rho ) \cdot \nabla i &=&  C  \nabla^2 i  + R_0 (\rho, r) s i   - i \label{eq:i4} \\
\frac{\partial r}{\partial t}   - C \nabla (\ln \rho ) \cdot \nabla r &=&  C  \nabla^2 r  + i \label{eq:r4}
\end{eqnarray}
We will first consider diffusion in one spatial dimension. \rededit{Before proceeding further, we note that  the SIR solution is obtained in the limit of large $C$, where from equations (\ref{eq:s4})-(\ref{eq:r4}) all spatial profiles become flat, and all variables become only functions of time, in a dependence which is identical to the SIR problem.}

\subsubsection{1-D geometries} 
\rededit{Assume finite values of $C$,} 1-D rectilinear geometries, and spatially constant density. The initial infection conditions are non-zero in a specified interval, \rededit{e.g.} $-1<x<1$, the rest of the domain being free of infections. We are interested in exploring how diffusion leads to the spreading of the contagion from the infected region, and particularly, whether or not traveling waves develop. 

Introducing a moving coordinate $\xi =x -V t $, where $V$ is the wave velocity, and assuming that a steady-state  (denoted by tilde) is reached in these coordinates one obtains,
\begin{eqnarray}
- V \frac{\partial \tilde{s}}{\partial \xi}   &=&   C  \frac{\partial^2 \tilde{s}}{\partial \xi^2}   - R_0  \tilde{s} \tilde{i}  \label{eq:swave} \\
 - V \frac{\partial \tilde{i}}{\partial \xi}  &=& C  \frac{\partial^2 \tilde{i}}{\partial \xi^2}  + R_0  \tilde{s} \tilde{i}  - \tilde{i} \label{eq:iwave} \\
  - V \frac{\partial \tilde{r}}{\partial \xi}  &=& C  \frac{\partial^2 \tilde{r}}{\partial \xi^2}  +  \tilde{i} \label{eq:rwave} 
\end{eqnarray}
subject to no-flux conditions at the ends
\begin{eqnarray}
\frac{\partial \tilde{s}}{\partial \xi} = \frac{\partial \tilde{i}}{\partial \xi} = \frac{\partial \tilde{r}}{\partial \xi} = 0, \mbox{ at } \xi = \pm \infty. 
\end{eqnarray}
The invariance of (\ref{eq:swave})-(\ref{eq:rwave}) to the transformation $\xi \to - \xi$,  $V \to - V$, suggests that there will be two asymptotic waves, one moving to the right, with velocity $V$, and one to the left, with the opposite velocity $-V$. Let $\xi_u$ and $\xi_d$ be two locations sufficiently upstream and downstream, respectively, of these two wave fronts. We then expect 
\begin{eqnarray}
\tilde{r}(\xi_u) = r_{V,\infty}, \tilde{i}(\xi_u) = 0, \mbox{ and } \tilde{r}(\xi_d) = 0, \tilde{i}(\xi_d) = 0 
\end{eqnarray}
where we have anticipated that $r_{V,\infty}$ might not be identical to the $r_\infty$ of the batch reactor result, equation (\ref{eq:rinfty}). 

The wave velocity can be determined by integrating (\ref{eq:rwave}) between $\xi_u$ and $\xi_d$,
\begin{eqnarray}
V = \frac{1}{r_{V,\infty}} \int_{\xi_u}^{\xi_d} \tilde{i} d \xi. \label{eq:wavevel1} 
\end{eqnarray}
and, equivalently,   
\begin{eqnarray}
V = \frac{1}{r_{V,\infty}} \int_{-\infty}^{\infty} \tilde{i} d \xi. \label{eq:wavevely} 
\end{eqnarray}
\rededit{We conclude that} a wave solution exists as long as the integral in (\ref{eq:wavevely}) is non-zero, which is always the case in a contagion. This wave spreads \rededit{at a constant velocity}, driven by diffusion, with a magnitude that expresses the intensity of the contagion.

\subsubsection{Invariance}
We can further explicitly remove the $C$-dependence by introducing re-scaled space coordinates and velocities, $\xi =\sqrt{C} \zeta$ and $V= W \sqrt{C}$. Denoted by superscript hat, the profiles satisfy
\begin{eqnarray}
- W \frac{\partial \hat{s}}{\partial \zeta}   &=&     \frac{\partial^2 \hat{s}}{\partial \zeta^2}   - R_0  \hat{s} \hat{i}  \label{eq:swave1} \\
 - W \frac{\partial \hat{i}}{\partial \zeta}  &=&   \frac{\partial^2 \hat{i}}{\partial \zeta^2}  + R_0  \hat{s} \hat{i}  - \hat{i} \label{eq:iwave1} \\
  - W \frac{\partial \hat{r}}{\partial \zeta}  &=&   \frac{\partial^2 \hat{r}}{\partial \zeta^2}  +  \hat{i} \label{eq:rwave1} 
\end{eqnarray}
subject to no-flux conditions at the two ends. The velocities becomes  
\begin{eqnarray}
W = \frac{1}{r_{V,\infty}} \int_{-\infty}^{\infty} \hat{i} d \zeta, \label{eq:wavevel2y} \\
\mathcal{V} = \frac{\sqrt{D \Lambda}}{r_{V,\infty}} \int_{-\infty}^{\infty} \hat{i} d \zeta. \label{eq:wavevel3y} 
\end{eqnarray}
Although the initially infected region $x \in (a,b)$ transforms into $\zeta \in (aC^{-1/2},b C^{-1/2})$, thus containing a $C$-dependence, the asymptotic traveling wave is independent of initial conditions, hence of $C$. \rededit{The resulting infection curves in the presence of diffusion as well as the wave velocity dependence on $R_{0}$ are further explored below}.  

\begin{figure}[h!]
    \centering
    \includegraphics[width=0.6\textwidth]{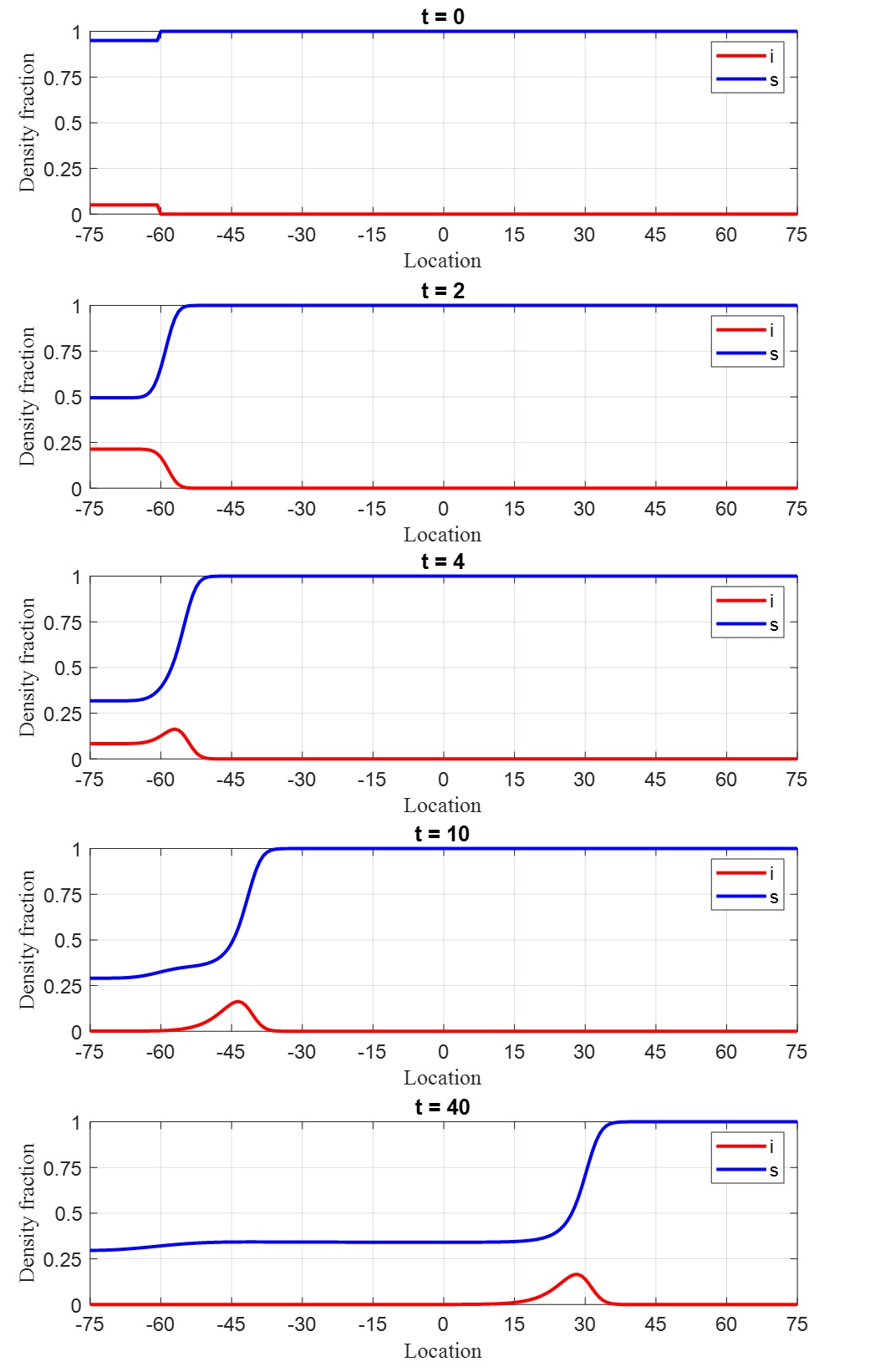}
    \caption{Infected and susceptible density fraction profiles at different values of time ($R_0(\rho,0) = 2.5$).}
    \label{fig09b}
\end{figure}

\begin{figure}[h!]
    \centering
    \includegraphics[width=0.6\textwidth]{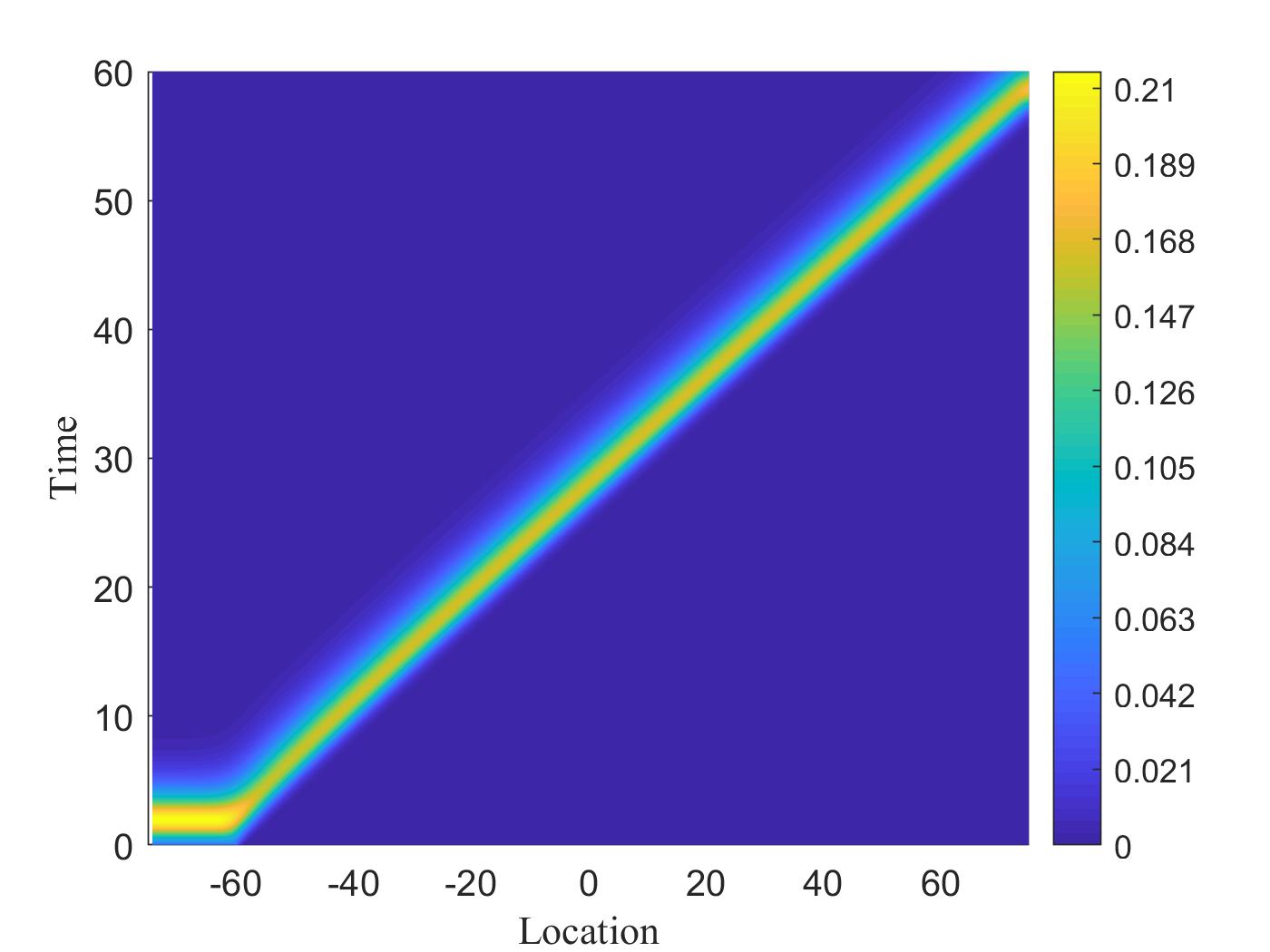}
    \caption{Infected density fraction as a function of space and time coordinates ($R_0(\rho,0) = 2.5$).}
    \label{fig09a}
\end{figure}

Numerical results of the solution of (\ref{eq:s4})-(\ref{eq:r4}) are shown in Figure \ref{fig09b}, for a problem in which the initial infection region is near the left boundary, for $R_{0}(\rho,0)=2.5$. As anticipated, the infection profiles evolve as a function of time, and reach an asymptotic traveling wave after about $t=3$. 
The wavelike nature of the solution is \rededit{evident}  if one plots the infected fraction in space-time coordinates (Figure \ref{fig09a}). A ridge with a constant slope is clearly seen. 

\begin{figure}[h!]
    \centering
    \includegraphics[width=0.6\textwidth]{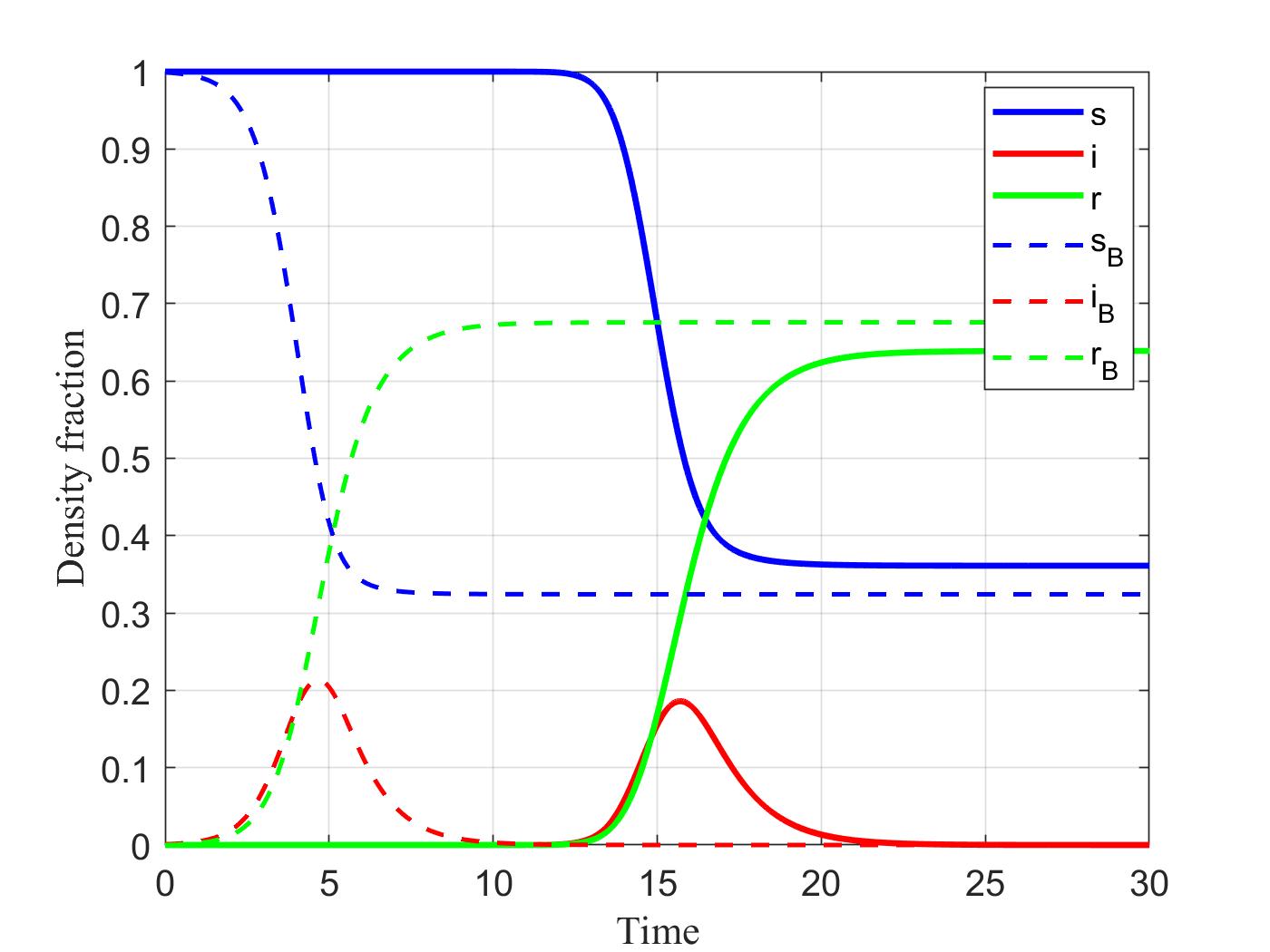}
    \caption{Variation of susceptible ($s$), infected ($i$) and recovered ($r$) fractions (steady state) as a function of non-dimensional time with diffusion included (solid lines) and for the batch reactor (SIR) problem  (dashed lines) for the same value of $R_0(\rho,0) = 2.5$. The initial infection fraction for the batch system is $i_B(0)=0.001$.}
    \label{fig10}
\end{figure}

\begin{figure}[h!]
    \centering
    \includegraphics[width=0.62\textwidth]{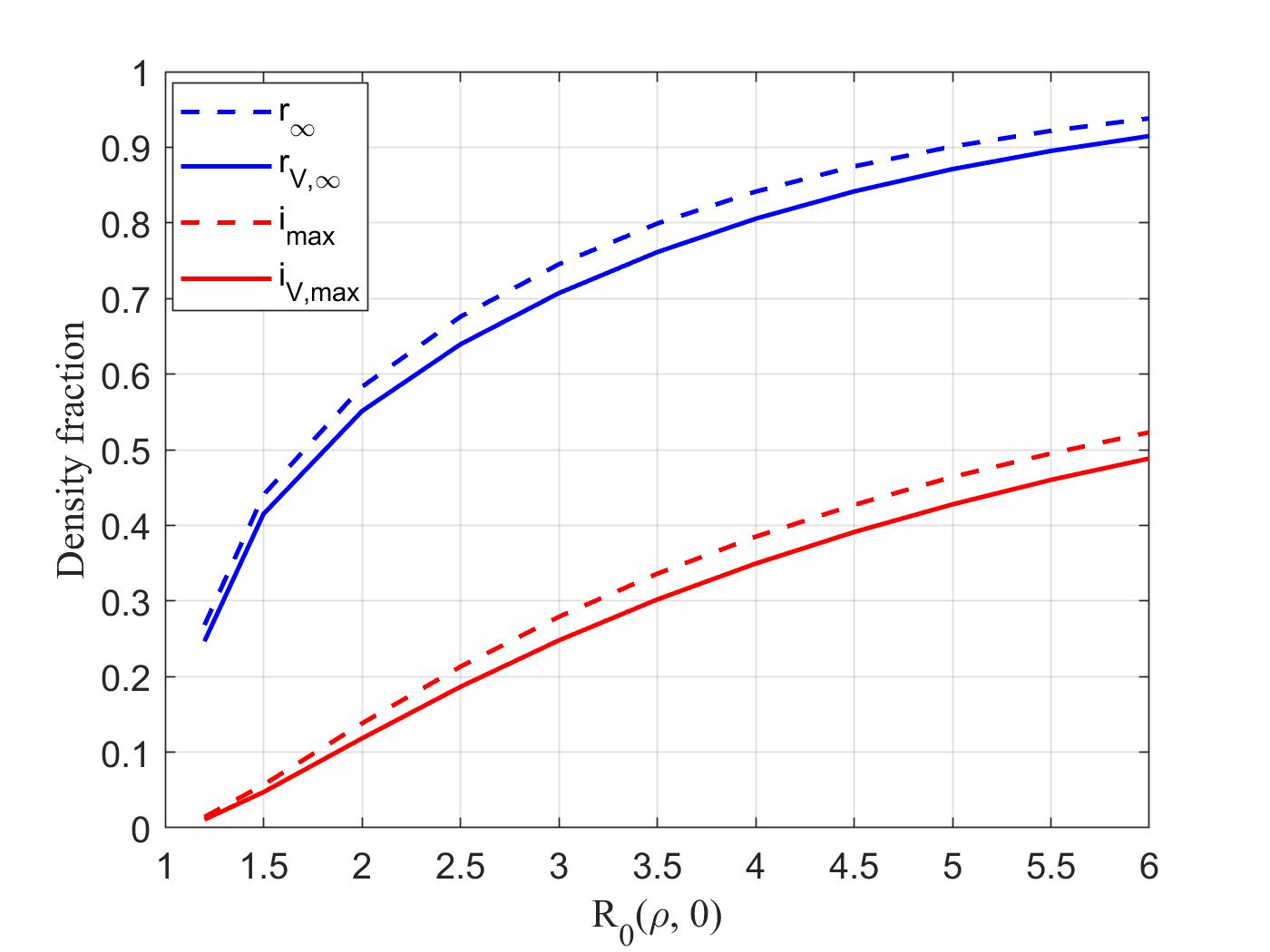}
    \caption{ Variation of herd immunity ($r_\infty$) and maximum infection fraction ($i_{max}$) as a function of $R_0(\rho,0)$ for the two cases of a batch system (dashed lines) and when diffusion is included (solid lines). The initial infection fraction for the batch system is $i_B(0)=0.001$.}
    \label{fig11}
\end{figure}

\rededit{Consider next, the question of how different are the infection profiles when diffusion is included}. Figure \ref{fig10} shows the relevant  wave profiles at a fixed value of $x$ (taken at  $x=-30$), calculated using the full system of equations, with diffusion included \rededit{and $C=...$}. Plotted also are the results of the batch reactor (SIR) problem, for the same value of $R_0(\rho,0)=...$. Diffusion does affect the shape of the curves obtained, leading to a slightly smaller (for this value of $R_0(\rho,0)$) infection intensity, essentially corresponding to a slightly smaller effective $R_0(\rho,0)$. Figure \ref{fig11} is a plot of the asymptotic value $r_{\infty}$ and of the maximum in infections $i_{max}$ as a function of $R_0(\rho,0)$ for the respective two cases. Diffusion acts to moderate somewhat the contagion intensity. \rededit{Overall,} the two solutions, corresponding to the contagion wave (equations (\ref{eq:s4})-(\ref{eq:i4})), now denoted by $i_D (t)$  and to the batch (SIR) problem (equations (\ref{eq:s2})-(\ref{eq:i2})), now denoted by $i_B (t)$, are approximately equal, \rededit{for this value of $R_0(\rho,0)$},  although they are not identical
\begin{eqnarray}
i(x,t) \to i(x-Vt) \equiv \hat{i}(\zeta) \equiv i_D (const - \frac{x}{V} + t) \approx i_B(const - \frac{x}{V} + t) \label{eq:wavevel3yy}
\end{eqnarray}
Note that the constant in (\ref{eq:wavevel3yy}) can absorbed in $x/V$.  

\begin{figure}[h!]
    \centering
    \includegraphics[width=0.6\textwidth]{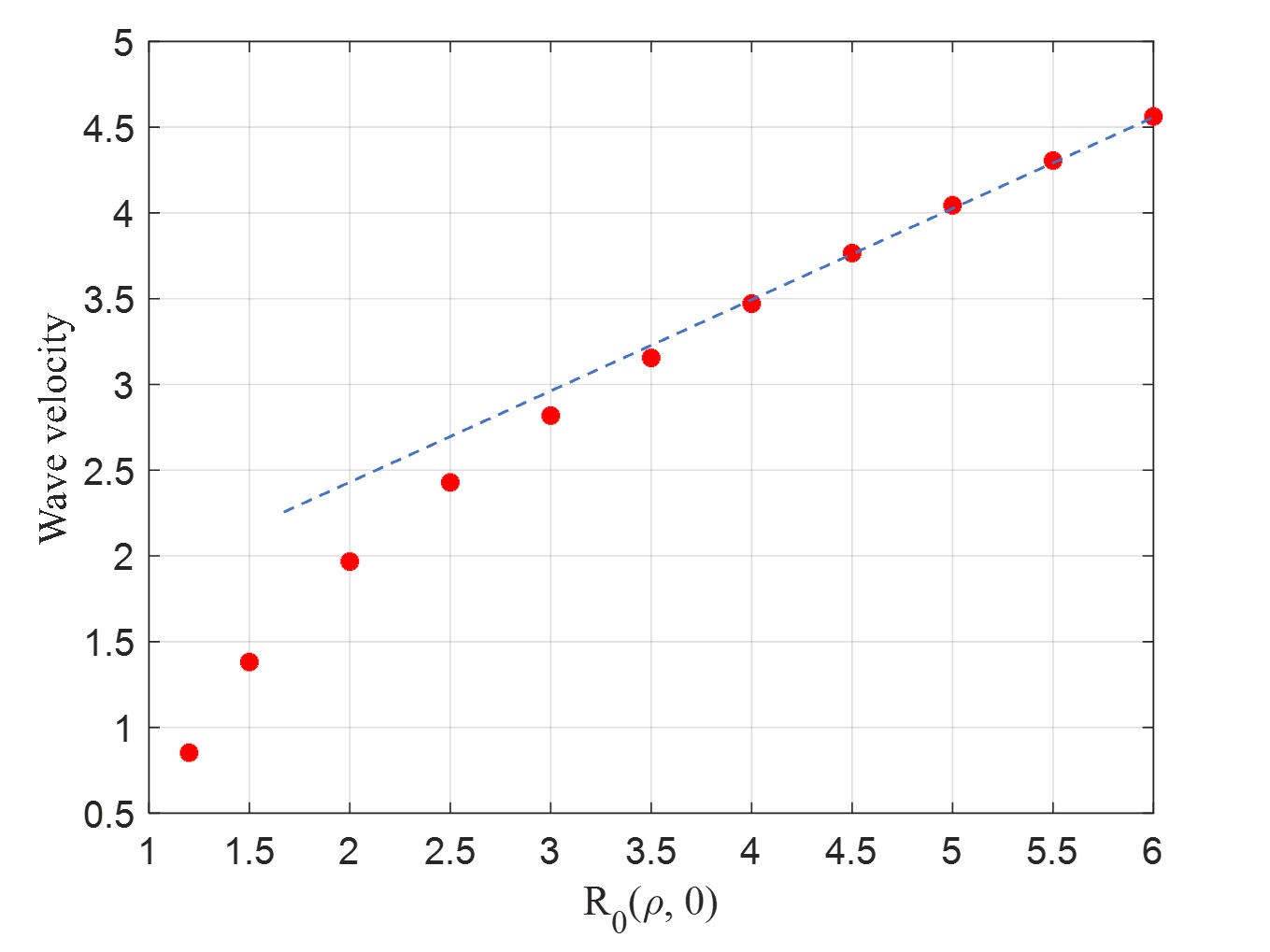}
    \caption{Variation of non-dimensional wave speed in a one-dimensional problem as a function of $R_0(\rho,0)$.}
    \label{fig12}
\end{figure}

The variation of the dimensionless velocity, equation (\ref{eq:wavevel2y}) with $R_0(\rho,0)$ is shown in Figure \ref{fig12}. As expected it increases with $R_0(\rho,0)$. Interestingly, it is  varying roughly in a linear fashion at relatively large values of the latter. 
\rededit{It is interesting to note that even for a finite value of the diffusion parameter $C$, the time profiles at a fixed spatial location are close to those for the SIR model, assuming sufficiently large $R_0(\rho,0)$. We can show this by introducing the new spatial variable $\chi = \zeta/W$ in equations (\ref{eq:s4})-(\ref{eq:r4}). For example, (\ref{eq:r4}) will read as}
\begin{eqnarray}
  - \frac{\partial \hat{r}}{\partial \chi}  &=&   \frac{1}{W^2}\frac{\partial^2 \hat{r}}{\partial \chi^2}  +  \hat{i} \label{eq:rwavechi} 
\end{eqnarray}
\rededit{Now, at relatively large values of $R_0(\rho,0)$, we can ignore the second derivative in the RHS of (\ref{eq:rwavechi}), which then becomes a set of equations identical to the SIR problem, subject to the equivalence $\chi \rightarrow -t$. This equivalence does not carry over when $R_0(\rho,0)$ is smaller and closer to 1, in which case the intensity of infection is smaller. Because this regime is of lesser interest, it will not be further explored here.} 

We conclude that, \rededit{ even in the absence of advection,   diffusion is sufficient to spread  the contagion at a constant velocity, with} the rate of spreading increasing with the square root of the diffusion coefficient and with the value of  $R_0(\rho,0)$. \rededit{Diffusion affects the overall infection intensity, although mildly, at larger $R_0(\rho,0)$ }. Restricting mobility, here represented by $D$, confirms a most important policy effect for containing infection. The finding that diffusion also affects somewhat the effective value of $R_0(\rho,0)$, is a result intuitively expected, but not previously identified. \rededit{It is also worth pointing out that because of the presence of the reaction terms, the effect of diffusion is to lead to fronts of constant velocity, as opposed to a velocity that varies with the square root of time, as is the case in the absence of reactions.}

\begin{figure}[h!]
    \centering
    \includegraphics[width=0.6\textwidth]{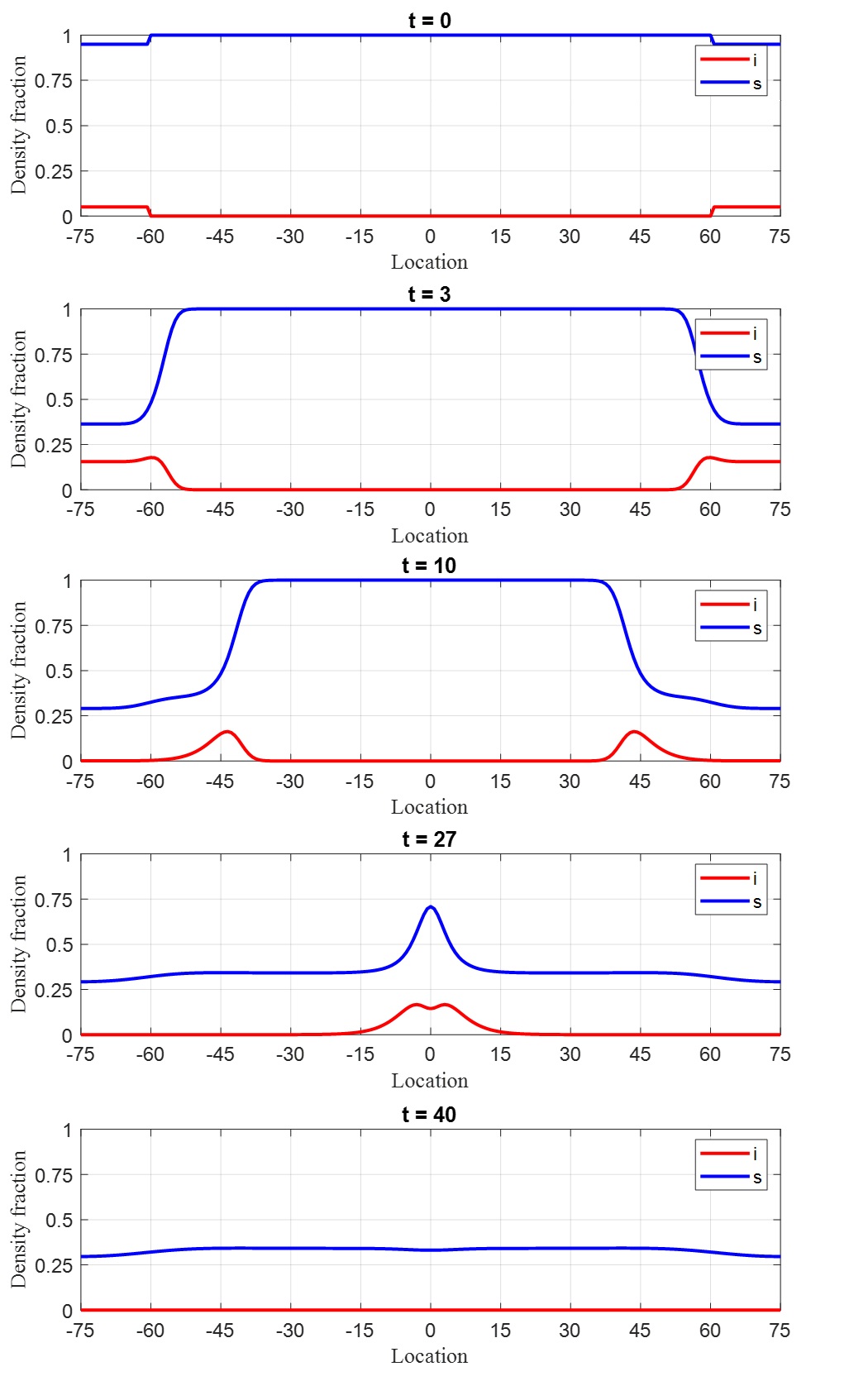}
    \caption{Infected and susceptible density fraction profiles at different values of time for two waves propagating towards the center of the domain ($R_0(\rho,0) = 2.5$).}
    \label{fig13b}
\end{figure}

For completeness, we simulated the “collision” of two waves, one moving from the left and the other from the right. Figure \ref{fig13b} shows how the two waves amplify as they interact, then ultimately decay as infection has spread fully and the population reaches conditions of herd immunity at the corresponding value of $R_0$. For a number of reasons, this wave is different from what is observed in other wave problems (e.g. solitary waves, where the wave velocities increase with the wave amplitude \cite{drazin1989solitons}). In the present context, the wave amplitude, e.g. $i_{max}$ cannot exceed the value of 1. 

\begin{figure}[h!]
    \centering
    \includegraphics[width=0.6\textwidth]{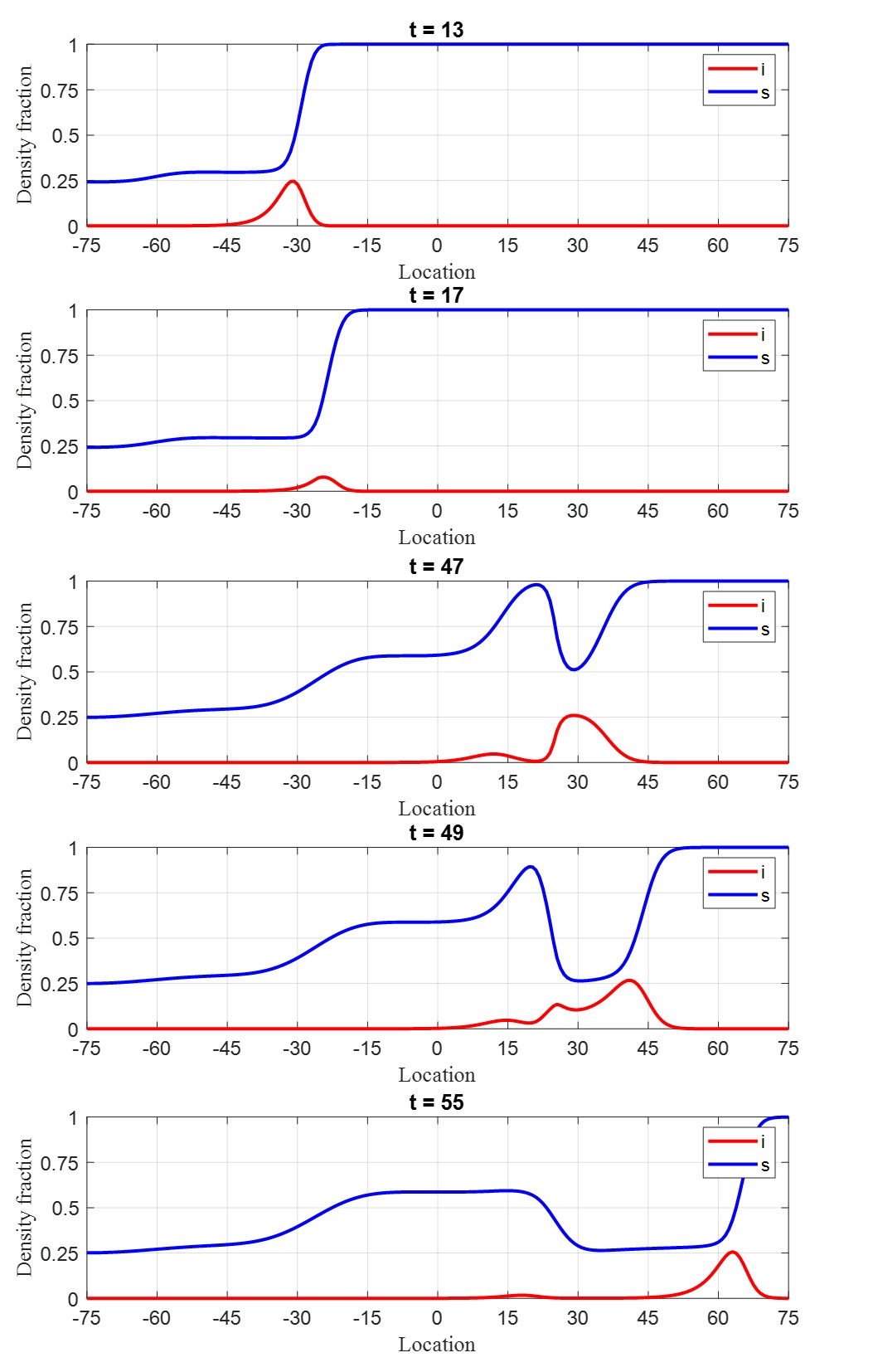}
    \caption{Infected and susceptible density fraction profiles at different values of time in a 1-D heterogeneous system. For $x \in (-80,-25)$ and $ x \in (25,80) $, $R_0(\rho,0) = 3$, while for $x \in (-25,25)$, $R_0(\rho,0) = 1.5$. }
    \label{fig15c}
\end{figure}

\begin{figure}[h!]
    \centering
    \includegraphics[width=0.6\textwidth]{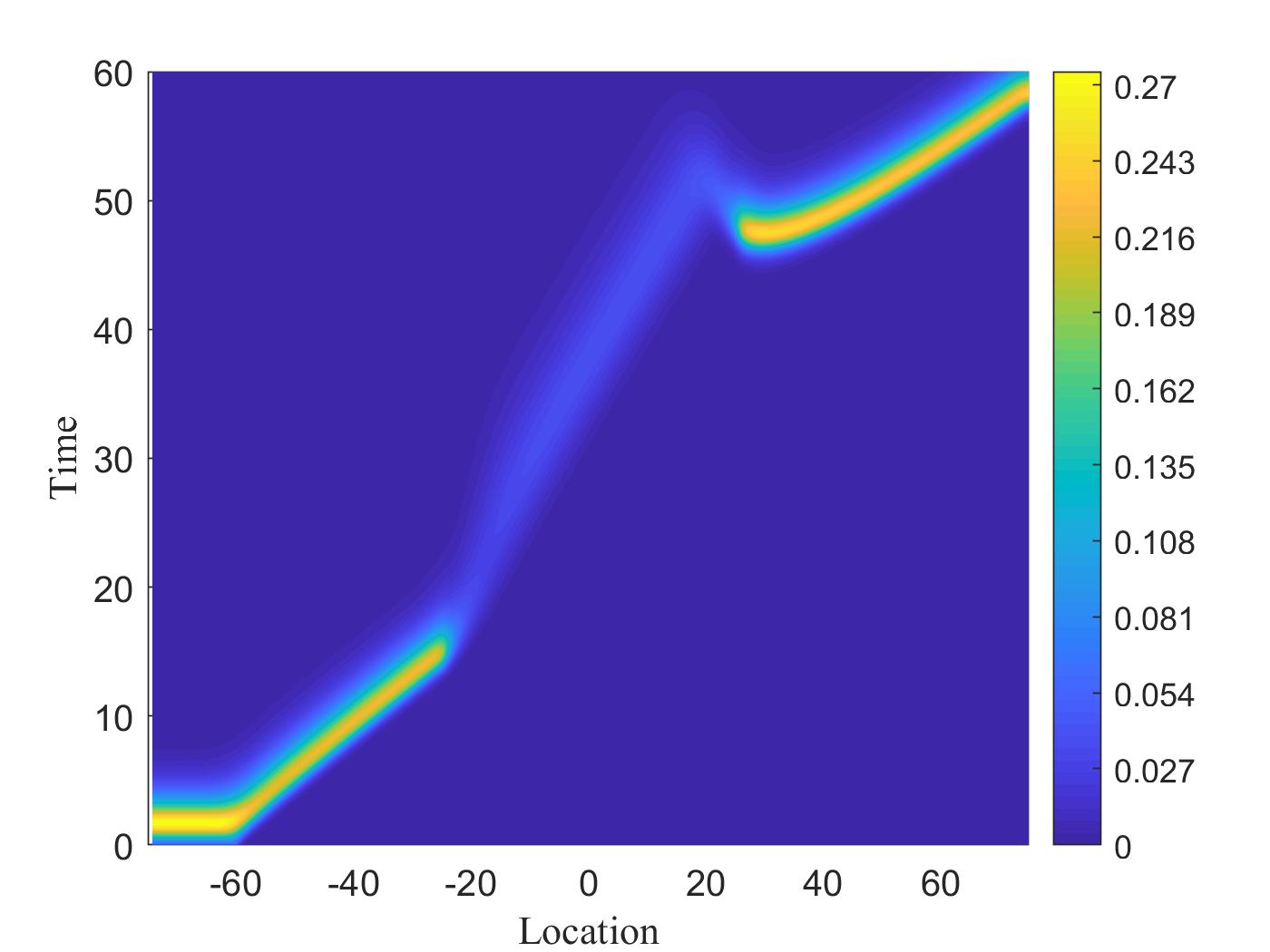}
    \caption{Infected density fraction as a function of space and time coordinates in a 1-D heterogeneous system: For $x \in (-80,-25)$ and $ x \in (25,80) $, $R_0(\rho,0) = 3$, while for $x \in (-25,25)$, $R_0(\rho,0) = 1.5$.}
    \label{fig15a}
\end{figure}

To conclude this section we next consider effects of heterogeneity in the 1-D rectilinear geometry. Figures \ref{fig15c} and \ref{fig15a} show results when an infection wave enters a region with a lower value of $R_0 (\rho,0)$, e.g. one of lower spatial density, from a region of a higher value of $R_0 (\rho,0)$, e.g. one of higher spatial density. For example, such could be the case of contagion spreading from an urban to a rural area. In the figure this occurs at $x = -25$. As it enters the low-density region the wave decelerates, and the magnitude of the infected fraction decreases. (The slowing of the wave is indicated by the increase of the slope in the $x-t$ domain.)  Conversely, at $x = 25$, when the wave now re-enters a region of higher $R_0(\rho,0)$, it accelerates to a larger velocity, the intensity of the infected fraction correspondingly increasing. 

\subsubsection{Radial Geometries}

\begin{figure}[h!]
    \centering
    \includegraphics[width=0.6\textwidth]{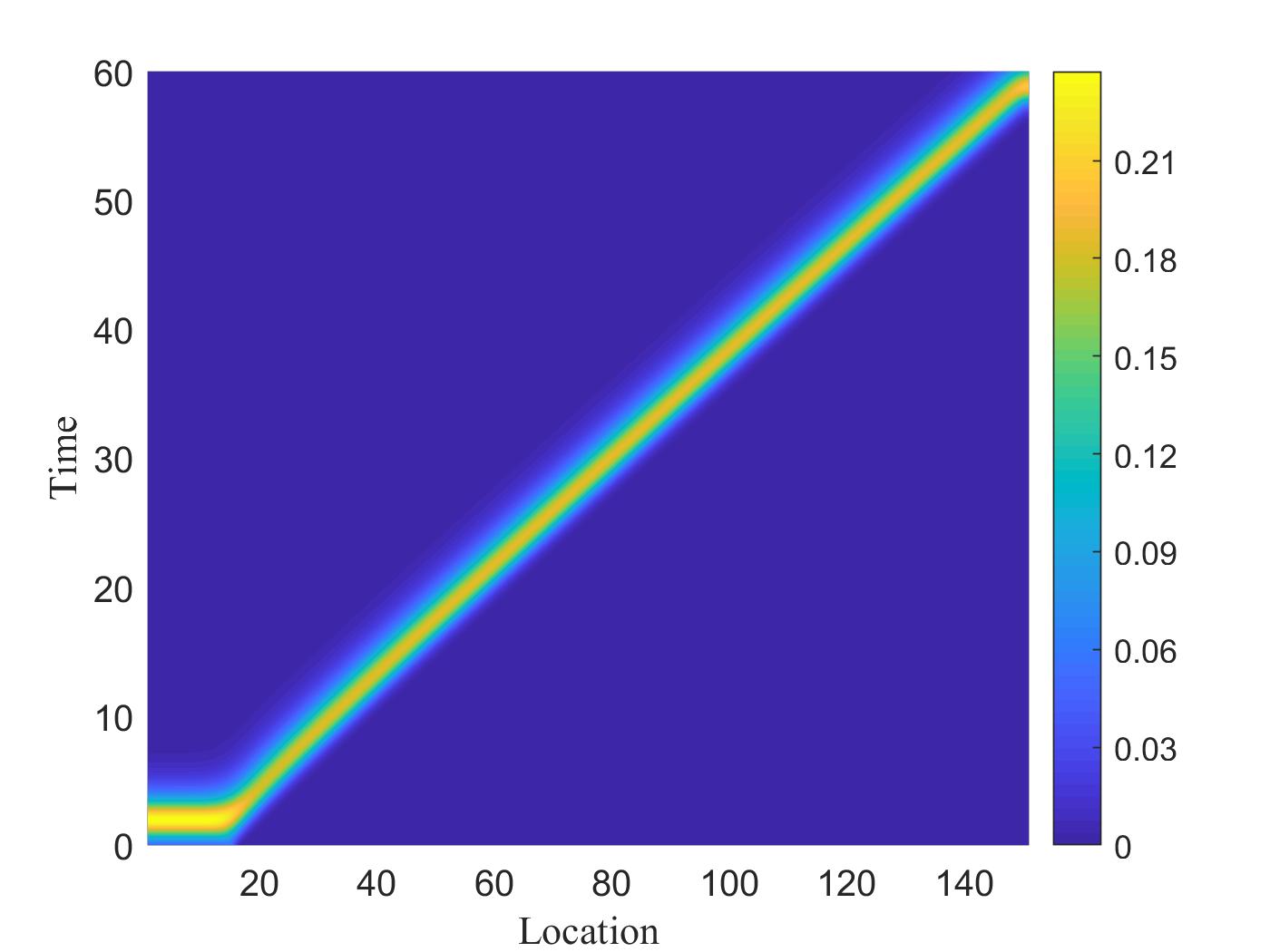}
    \caption{Infected density fraction as a function of radial (space) and time coordinates ($R_0(\rho,0) = 2.5$).}
    \label{fig14}
\end{figure}

The results of the 1-D rectilinear geometry apply almost identically to radial geometries. \rededit{Consider first extension to radially symmetric 2-D geometries. An extension to more general 2-D geometries is considered in the subsequent section.} Now, equations (\ref{eq:s4})-(\ref{eq:i4}) become
\begin{eqnarray}
\frac{\partial s}{\partial t}    &=&   \frac{C}{\mu}  \frac{\partial }{\partial \mu }( \mu \frac{\partial s }{\partial \mu })   -R_0 (\rho, r) s i  \label{eq:s5} \\
\frac{\partial i}{\partial t}  &=& \frac{C}{\mu}  \frac{\partial }{\partial \mu }( \mu \frac{\partial i }{\partial \mu })  + R_0 (\rho, r) s i   - i \label{eq:i5} \\
\frac{\partial r}{\partial t}    &=&  \frac{C}{\mu}  \frac{\partial }{\partial \mu }( \mu \frac{\partial r }{\partial \mu })  + i \label{eq:r5}
\end{eqnarray}
where we used $\mu$ to denote the radial coordinate.  We look for the solution of the problem when a region around the origin ($\mu=0$) is initially infected. As in the rectilinear geometry case, we expect that the solution will evolve in terms of a traveling wave, therefore, we consider a transformation to the moving coordinates $\xi = \mu - Vt $, and $t' = t$.
In these coordinates, the equations become 
\begin{eqnarray}
\frac{\partial s}{\partial t'} - V \frac{\partial s}{\partial \xi}     &=&  C \frac{\partial^2 s }{\partial \xi^2 } +  \frac{C}{\xi + Vt'}  \frac{\partial s }{\partial \xi }   -R_0 (\rho, r) s i  \label{eq:s6} \\
\frac{\partial i}{\partial t'} - V \frac{\partial i}{\partial \xi}     &=&  C \frac{\partial^2 i }{\partial \xi^2 } +  \frac{C}{\xi + Vt'}  \frac{\partial i }{\partial \xi }  + R_0 (\rho, r) s i   - i \label{eq:i6} \\
\frac{\partial r}{\partial t'} - V \frac{\partial r}{\partial \xi}     &=&  C \frac{\partial^2 r }{\partial \xi^2 } +  \frac{C}{\xi + Vt'}  \frac{\partial r }{\partial \xi } + i \label{eq:r6}
\end{eqnarray}
\rededit{By further taking the limit of large $t'$}, equations  (\ref{eq:s6})-(\ref{eq:i6}) transform to the same equations as (\ref{eq:swave})-(\ref{eq:iwave}). \rededit{It follows that} the same results  hold for radial geometries as for the rectilinear problem. The simulations for the 1-D radial geometry, shown in Figure \ref{fig14}, confirm these findings. 
\rededit{As before, the asymptotic solution in this case of reaction-diffusion system} does not enter in the familiar form of a similarity variable, $\eta=\frac{\mu}{\sqrt{t}}$, but rather in terms of a wave that propagates with a constant linear speed.  

\subsubsection{Effect of Heterogeneity in Two Dimensions}
Consider, now \rededit{how infection propagates in a general, heterogeneous 2-D system, via diffusion. We are interested in understanding how propagation occurs and whether or not the asymptotic wave solutions obtained for the 1-D geometries apply here as well.} In particular, we are further interested in knowing if one can use a wave equation to describe the evolution of the contagion fronts. To explore this question we consider two different geometries, one corresponding to a layered (stratified) system, and one corresponding to growth in a heterogeneous system.  

\begin{figure}[h!]
    \centering
    \includegraphics[width=0.7\textwidth]{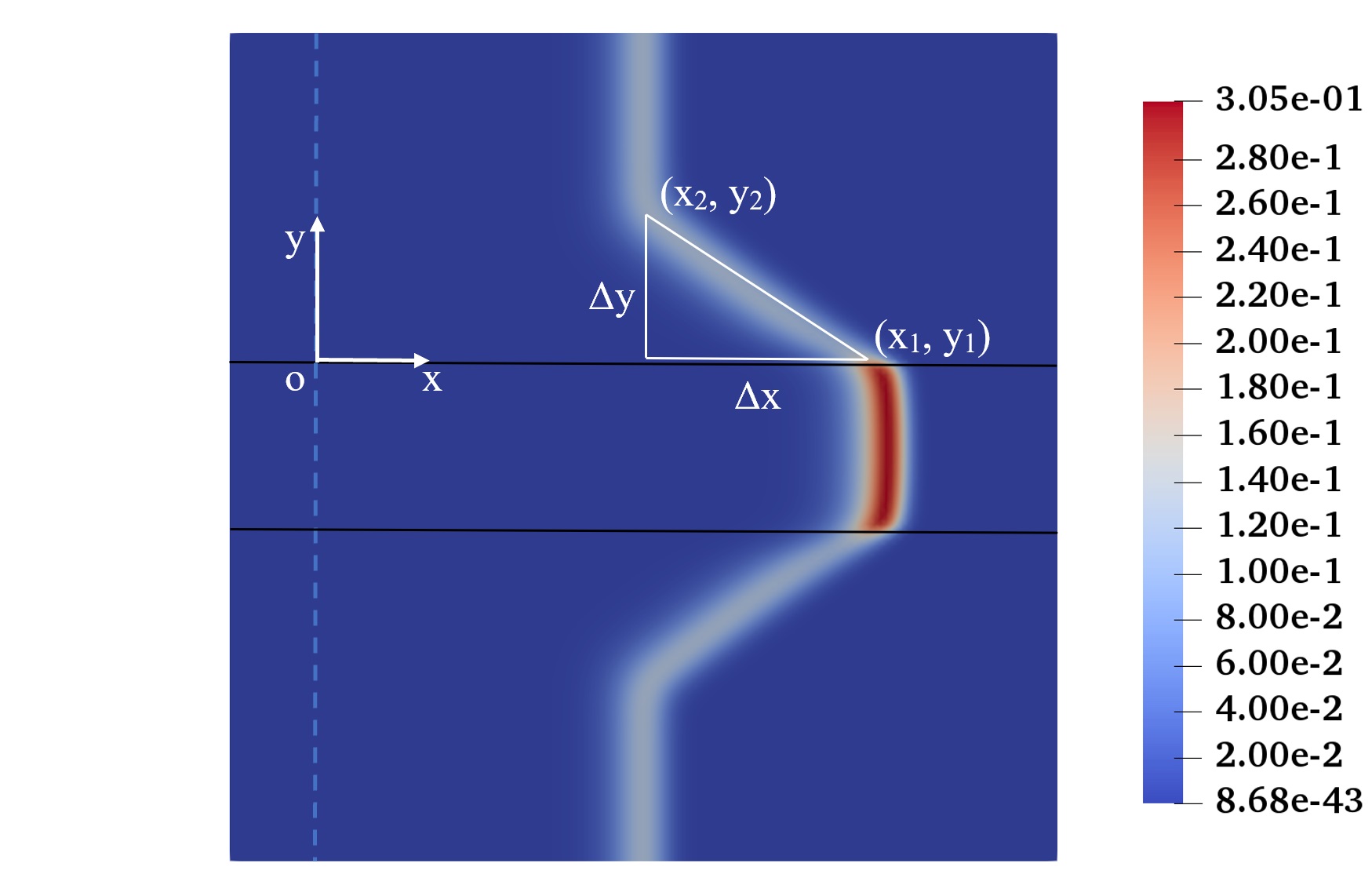}
    \caption{Infected density fraction ($i$) at $t=40$ in a layered medium with an infection wave starting at $x = 0$. In the outer layers $R_0(\rho,0) = 2$ while in the inner layer $R_0(\rho,0) = 3.5$. The spatial dimension of the domain is $150\times150$.}
    \label{fig17}
\end{figure}

\paragraph{Layered System} 
Consider propagation in a stratified layered system, the middle layer having a larger density value (thus, a larger $R_0 (\rho,0)$), compared to its two adjacent layers. Infection is initiated uniformly on the left boundary (see Figure \ref{fig17}). 
As expected, an infection wave emerges in the middle layer, traveling with a velocity $v_1$ corresponding to the high $R_0$ value in that layer. \rededit{ This is indeed equal to the corresponding asymptotic 1-D wave velocity  of 3.15} (see Figure \ref{fig12} for $R_0(\rho,0) = 3.5$). A slower moving contagion front develops in the adjacent layers, with corresponding velocity $v_2$ equal to 1.96, \rededit{which is also } the asymptotic value of the 1-D wave velocity  corresponding to $R_0(\rho,0) = 2$. Because of diffusion, however, transverse waves also emerge, emanating at the interface of the layers, propagating in the transverse ($y$) direction. These waves are manifested as linear fronts, slanted with respect to the transverse direction, but moving forward with speed $v_1$. 

If we set the origin of the coordinates at the intersection of the interface between the two layers and the initial infection boundary, the straight line connecting the two fronts in the two layers is described by the following equation
\begin{eqnarray}
 F\equiv y(v_1- v_2)+x a -v_1 a t =0 \label{eq:Frontx}
\end{eqnarray}
where the leading edge travels with the middle layer velocity $v_1$, the trailing edge with the adjacent layer velocity $v_2$, and the tip of the connecting layer at $y=y_2(t)$ with velocity $a$, to be determined. Because the waves emanating from the interface travel in the adjacent layers at fixed $x$ with transverse velocity $v_2$, equation (\ref{eq:Frontx}) gives
\begin{eqnarray}
 v_2(v_1- v_2)=v_1 a \label{eq:Front7}
\end{eqnarray}
therefore we find
\begin{eqnarray}
a= \frac{v_2}{v_1}(v_1- v_2)\label{eq:Front8}
\end{eqnarray}
The slope of the straight line must then be
\begin{eqnarray}
slope= \frac{a}{(v_1-v_2)}=\frac{v_2}{v_1} \label{eq:Front9}
\end{eqnarray}
The simulations of (\ref{fig17}) show that this is indeed the case. 

More generally, if we define a front by $F(\bm{x},t)=0$ where $\bm{x}$ is the space vector, its evolution satisfies 
\begin{eqnarray}
 F_{t}(\bm{x},t)+\bm{v}\cdot \nabla F&=&0    \label{eq:Front2} \\
  F_{t} (\bm{x},t)+v_n |\nabla F|&=&0 \label{eq:Front3}
\end{eqnarray}
where subscript $t$ denotes time derivative, and $v_n$ is its component in the direction of $\nabla F$. For rectilinear or radial geometries, as well as in each of the two layers  discussed above, (\ref{eq:Front3}) reduces to a linear wave propagating at a constant velocity. However, for the straight line connecting the two waves,  resulting from a sequence of waves emanating from the layer interface, the relevant equation is (\ref{eq:Frontx}), which leads to a normal velocity equal to 
\begin{eqnarray}
v_n = \frac{v_2}{\sqrt{(1+\frac{v_2^2}{v_1^2})}} \label{eq:Front10}
\end{eqnarray}
The interdependence between the three velocities (leading front, $v_1$, normal to the surface, $v$, and trailing front, $v_2$) is to be noted.   

\begin{figure}[h!]
    \centering
    \includegraphics[width=0.8\textwidth]{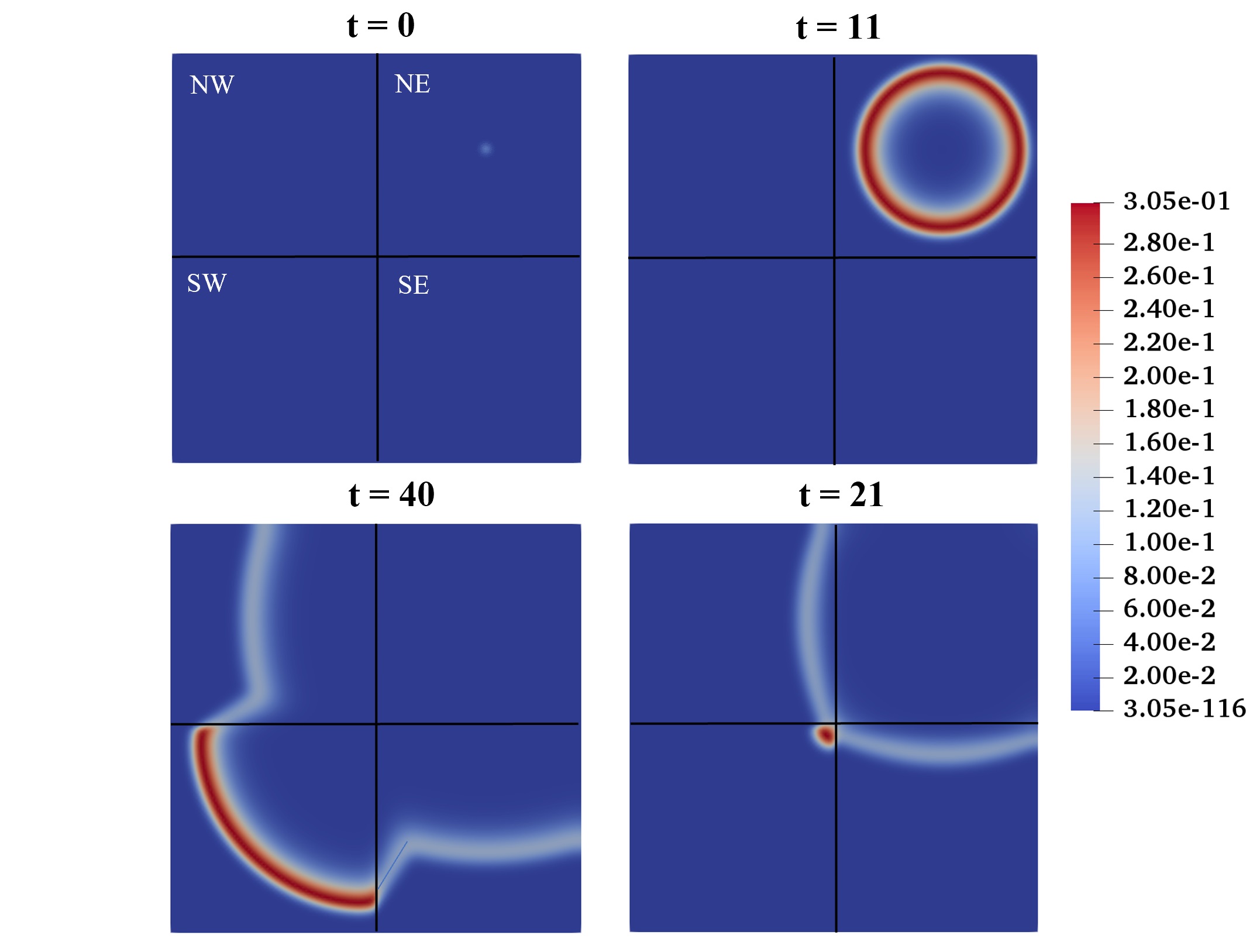}
    \caption{Contour plot of the infected density fraction a region with four quadrants with different values of $R_0(\rho,0)$ at different values of time. $R_0(\rho,0)=3.5$ for NE and SW, and $R_0(\rho,0)=2$ for NW and SE, respectively. The spatial dimension of the region is $150\times150$.}
    \label{fig16}
\end{figure}

\paragraph{Four heterogeneous systems} A different illustration of these effects is shown in Figure \ref{fig16}, where infection initiates in the upper right corner of a rectangular domain with four different quadrants (NE, SE, NW and SW), each taken with a different density (hence different values of $R_0 (\rho,0)$). \rededit{The idea is to simulate how infection can spread in a "country" consisting of four hypothetical quadrants, with different densities, hence different values of $R_0 (\rho,0)$}. Indeed, a simulation of the spreading of infection in the Lombard region of Italy using a similar continuum model was provided in \cite{viguerie2020diffusion}. In our work, we have assumed that the two diagonally opposite quadrants (e.g. NE and SW) have relatively larger values of $R_0 (\rho,0)=3.5$, hence an asymptotic wave velocity of 3.15, the other two (SE, NW) being of relatively smaller values of $R_0 (\rho,0)=2$, and with an asymptotic wave velocity of about 2. The infection waves follow the expected pattern. Infection grows first radially in the NE quadrant with the asymptotic speed of 3.15 (upper right panel in Figure \ref{fig16}). When the SE and NW boundaries are encountered (lower right panel in Figure \ref{fig16}), the waves slow down and start spreading in a radial manner as they enter the two quadrants at the slower speed of 2. Subsequently, the infection wave enters the high density SW region, in which it starts spreading radially, now moving with the higher velocity of 3.15 (lower left panel in Figure \ref{fig16}). Upon touching the boundaries with the NW and SE regions, diffusion causes infection waves to start emanating from the higher to the lower density regions, resulting into a flat front, similar to that for the layered system, which connects the two waves in the two different regions. For the same reasoning as in the layered system, this front has a slope equal to the ratio of the two respective velocities, namely (\ref{eq:Front9}). The simulations confirm these results. \rededit{Similar results can be obtained when the four quadrants are heterogeneous, where the values of $R_0 (\rho,0)$ are distributed in a stochastic manner in the regions. 3}   

We end this section by noting that one can further explore the use of a wave (or eikonal) equation for the description of the spreading of contagion. This would be the objective of a future study. 

\rededit{\subsection{Effect of Advection}}

\rededit{Consider, next, the effect of advection. For simplicity, we proceed by 
keeping the same advective velocity for all three populations. The corresponding dimensionless equations read as follows
\begin{eqnarray}
\frac{\partial s}{\partial t} + Da \bm{v}\cdot\nabla s &=& C\nabla^2 s-R_0 (\rho, r) s i  \label{eq:ad1} \\
\frac{\partial i}{\partial t} + Da \bm{v}\cdot \nabla i &=& C\nabla^2 i + R_0 (\rho, r) s i   - i \label{eq:ad2} \\
\frac{\partial r}{\partial t} + Da \bm{v}\cdot\nabla r  &=&  C\nabla^2 r+ i \label{eq:ad3} 
\end{eqnarray}
along with (\ref{eq:density_nondim}). The simplest result is obtained when one assumes that the velocity is constant in space and time. Consider the case of 1-D geometries, and take without loss $\bm{v}=\bm{e}_x$, where $bm{e}_x$ is the unit vector in the $x$-direction. By following the same reasoning as the case of diffusion, we can readily arrive at the following results corresponding to the travelling waves, equation (\ref{eq:wavevel2y}). We find that in the forward moving wave the velocity is enhanced
\begin{eqnarray}
W = v + \frac{1}{r_{V,\infty}} \int_{-\infty}^{\infty} \hat{i} d \zeta, 
\label{eq:waveveadvl2y} 
\end{eqnarray}
where we defined $v=Da/\sqrt{C}$. 
Conversely, the backwards moving wave is retarded leading to a velocity of 
\begin{eqnarray}
W = -v + \frac{1}{r_{V,\infty}} \int_{-\infty}^{\infty} \hat{i} d \zeta
\label{eq:wavevebackl2y} 
\end{eqnarray}
As expected the contribution of advection in this constant advection case is to speed up (or slow down) in the direction of advection the spreading velocity with a contribution that is equal to $Da/\sqrt{C}= U/\sqrt{D\Lambda}$.}

 \begin{figure}[h!]
    \centering
    \includegraphics[width=0.8\textwidth]{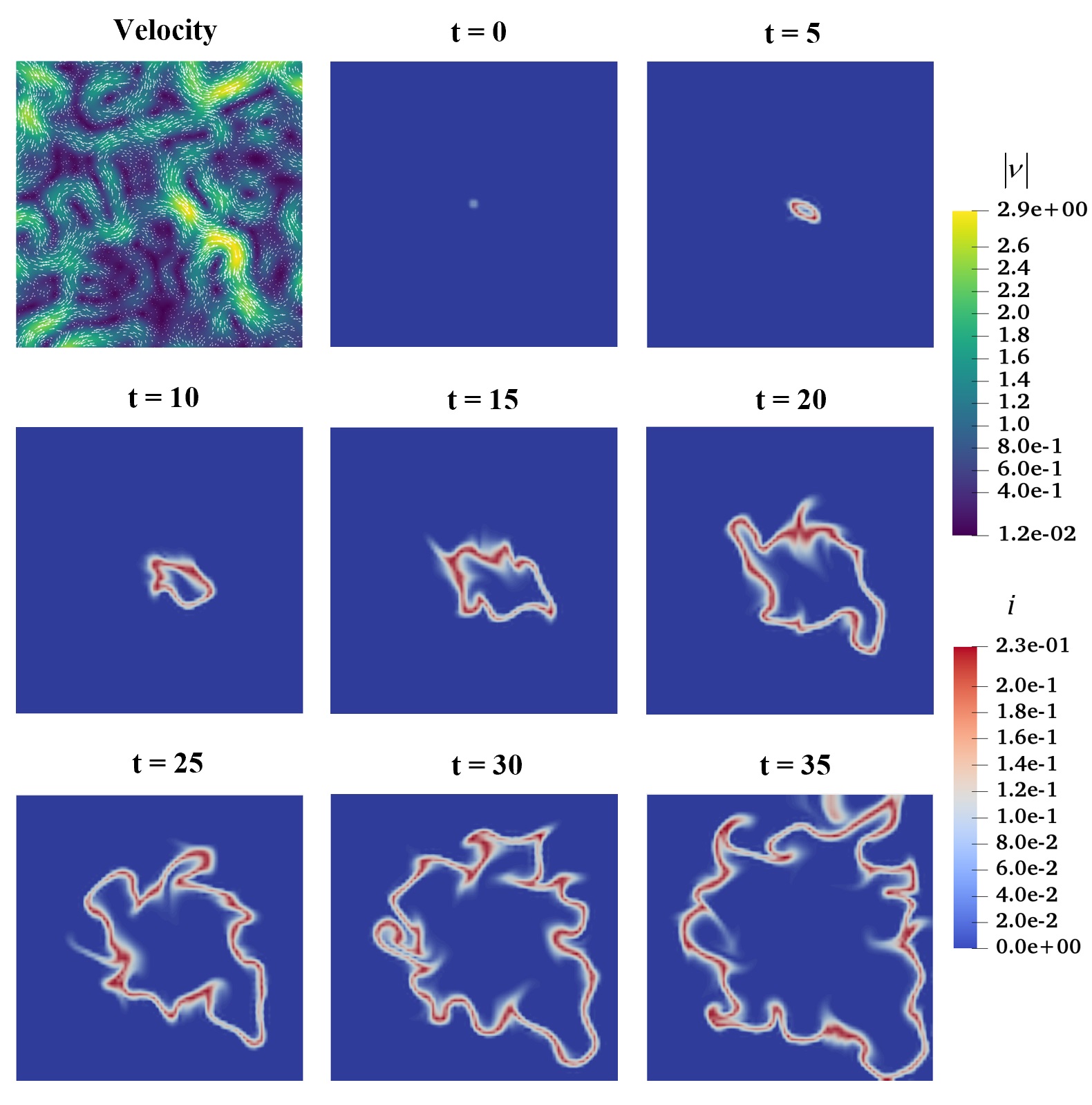}
    \caption{Simulation in a domain with random velocity field and constant diffusion and $R_0$. Top left: velocity field for the simulation. From top left to bottom right: images of the infected fraction at different time instances.}
    \label{figadvection}
\end{figure}

\rededit{More interesting is the problem when the velocity vector is a stochastic variable, distributed heterogeneously in space. Such a distribution will render all species to be stochastic variables as well and all relevant variables become ensemble averages. We have encountered some of this feature in a previous section dealing with fluctuations. The effect of heterogeneity will enter in two ways: as a correction to the non-linear reaction terms, via the ensemble average of the product of fluctuations, as discussed previously; and as the ensemble average of the product of the fluctuations of the velocity with the gradient of the fluctuation, namely $<\bm{v}^{\prime}\cdot\nabla s^{\prime}>$. The latter effect arises in many similar contexts, for example in the advection of a scalar in turbulent flow \cite{sreenivasan2019turbulent}, or in transport in heterogeneous porous media \cite{gelhar1983three}. It will manifest itself in the form of an effective macrodispersivity, which is dominant over molecular diffusion, namely 
\begin{eqnarray}
<\bm{v}^{\prime}\cdot\nabla s^{\prime}>= D_{macro}\nabla^2<s> \label{eq:Macro}
\end{eqnarray}
and likewise for the other species. For example, in the case of flow in heterogeneous porous media, the macrodispersivity is proportional to the mean velocity and to a characteristic correlation length $\lambda$,
\begin{eqnarray}
D_{macro}= \alpha|\bm{v}|\lambda \label{eq:Macro2}.
\end{eqnarray}
}

\rededit{We have simulated the solution of a 2-D problem, with $Da=1$, $C=0.1$, and where the velocity field was distributed stochastically in a divergence-free field, with zero mean, a dimensionless correlation length of 0.05, and $R_0(\rho, 0)= 2.5$. Figure \ref{figadvection} shows in one of the panels the velocity field. The remaining panels show state of the infected fraction at different time instances. We note that the wave propagates approximately radially; however the front bends and folds at several places that are determined by the local velocity field.}

 \begin{figure}[h!]
    \centering
    \includegraphics[width=0.8\textwidth]{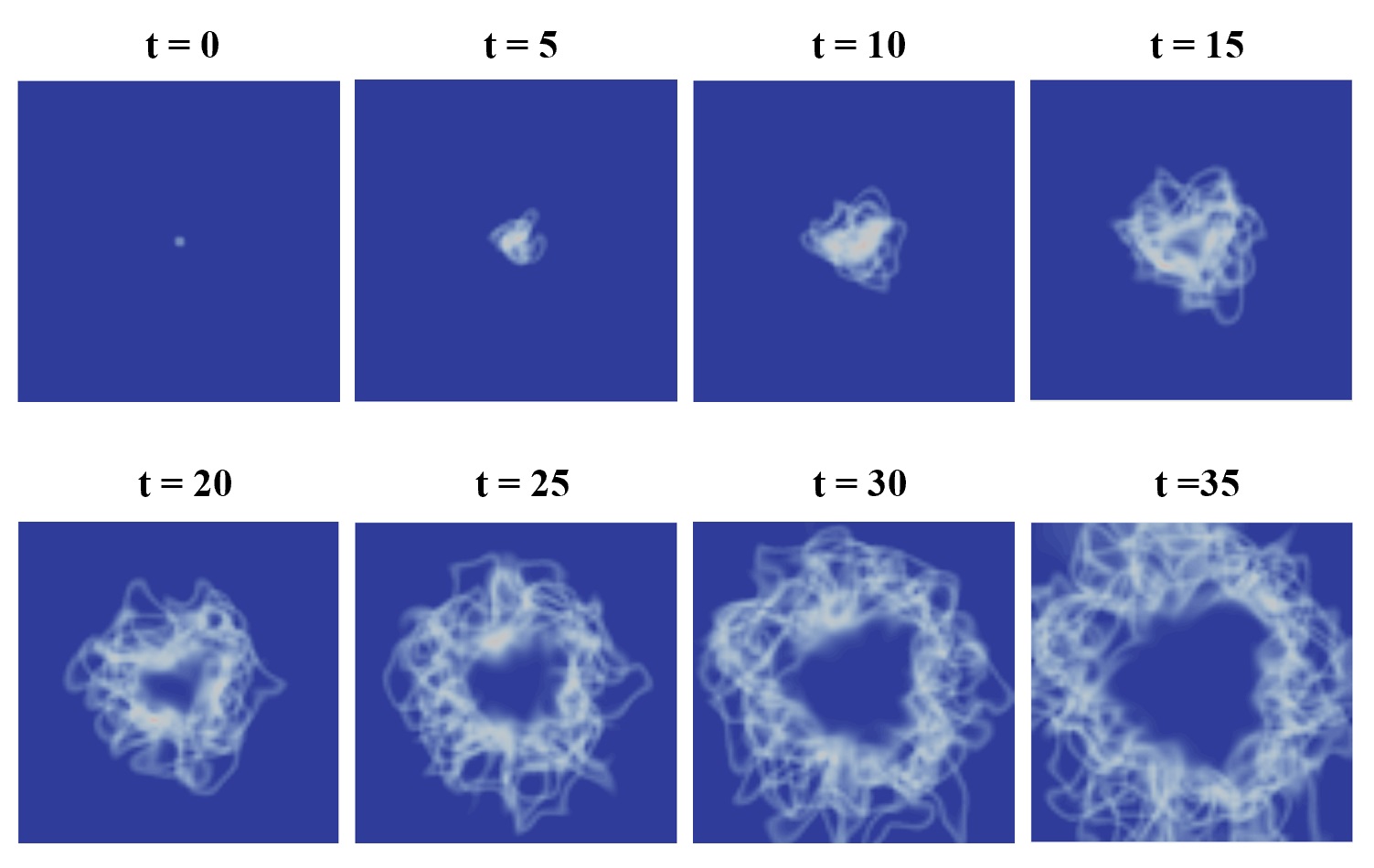}
    \caption{Simulation in ten different domains with random velocity field and constant diffusion and $R_0$. From top left to bottom right: ensemble-averaged images of the infected fraction at different time instances.}
    \label{figensemble}
\end{figure}

\rededit{In Figure \ref{figensemble} we have plotted the ensemble average of the infection front obtained from ten such random velocity fields. The results suggest a wave that moves radially at a constant speed, reminiscent of the spreading in the radial geometry driven by diffusion alone. The thickness of the front is substantially larger, however, indicating the enhanced dispersion, as anticipated. More quantitative analyses of this important effect will be the subject of a future communication.}

\rededit{\subsection{Additional Remarks}}

The wavelike spread of infections predicted by the model has been observed in several pandemics. In this context it is interesting to compare the spread of the 1347-1350 black death (pneumonic plague) in Europe and that of the 2009 pandemic influenza in the US. While there is significant debate regarding the effective $R_0$ value for these pandemics, it is generally accepted that they are in the range of $R_0 \approx 1.5-2.5$ \cite{noble1974geographic,nikbakht2019comparison}. Further, for both diseases the infectious period is around two weeks, and therefore $\Lambda \approx 1/14 \; days^{-1}$. To predict the wave speed, we need an estimate of the respective diffusion (or dispersion) coefficients. It is to be expected that in the medieval times,  the mobility of humans, therefore of infection, was much smaller. In \cite{noble1974geographic,rashevsky1968looking} this was determined by how quickly information was disseminated (in that case by word of mouth), resulting into an estimate of about $27 miles^2/day$. \rededit{This can be interpreted as either a diffusivity or a macrodispersivity.} 
Using the obtained estimate in our estimate for the speed of the contagion (see Figure \ref{fig12}) we arrive at a speed of $2.7 \; miles/day$ for the black death pandemic, which is in the ball-park of the actual observed speed of around $1 \; mile/day$ \cite{langer1964black}.

For the 2009 influenza pandemic, we can make a similar back-of-the-envelope calculation. 
Because of the much higher increased mobility, and the lack of significant public health awareness, we can estimate that the equivalent random walk radius was an order of magnitude larger compared to the medieval times. We will assume a diffusion coefficient of  $2,500 \; miles^2 day^{-1}$, or equivalently, a diffusivity of about $10^5 m^2/sec$. Clearly, the mean radius of $50 miles$ so assumed incorporates a mix of various travel or other commute activities in the modern world. 
Since children play an important role in the spread of influenza, we derive this estimate based on their activities. We assume that a typical child will see 30 other children in school and playgrounds during a typical day. Further, they will travel an average of 10 miles to get to these places. In a coarse analogy to an ideal gas this gives a mean-free path of 10 miles and a frequency of collision of $30 \; days^{-1}$, and therefore a diffusivity \rededit{(or macrodispersivity)} of around $3,000 miles^2/day $.   
Using these parameters in our estimate of the speed of the contagion wave, we arrive at a speed of $29 \; miles/day$ for the 2009 influenza pandemic. This is also in the ball-park of the observed wave speed of around $23 \; miles/day$ for the 2009 influenza pandemic \cite{gog2014spatial}. The higher diffusion/dispersion coefficient for the 2009 influenza  pandemic leads to a significantly faster wave speed. \rededit{All these results should be also interpreted based on the understanding that the enhanced diffusion coefficient reflects in all likelihood macrodispersion driven by the fluctuating advective velocity field. Further work in this area is needed.}   

\section{Conclusions}

In this paper we used an analogy with chemical reaction engineering processes to model the growth and spreading of human-transmitted infections, such as COVID-19 and influenza. The basis of the model is the assumption of three distinct populations, as in the celebrated SIR model, which are mapped into equivalent chemical species. An important first result from this formulation is that the relevant quantities are population densities (specifically, areal spatial densities) and not total populations, as is in the SIR model. These satisfy partial differential equations, with spatial mobilities included through diffusion and advection. Spatial density effects are then incorporated into the kinetic constant of the infection growth, hence into the key parameter $R_0$, which becomes an explicitly function of spatial densities, and found to decrease with the extent of the contagion. Incorporating a spatial density dependence in contagion kinetics is necessary, and consistent with health guidelines and droplet dynamics.  

In the \rededit{case of perfect mixing, where profiles are spatially uniform}, the results revert to a modified version of the SIR model, in a geometry recognized as a “batch reactor”. We find that if were to identify an effective SIR-like $R_0$, it  would be smaller than when spatial effects \rededit{(namely, density effects) on the kinetic coefficient} are ignored, which is the standard SIR case. The infection curves are found to be largely independent of initial conditions, the effect of which is simply to delay the onset of the infection. Analogous results hold when a ban on “imported infection” is applied, which also only affects to delay the onset of the epidemic, assuming that $R_0$ is not modified. Small fluctuations in the initial conditions have a minimal effect on the ensemble behavior. However, this is not the case when the fluctuations are significant in space or in time, in which case they result into non-trivial ensemble averages. Using the density-dependent $R_0$ we can readily model the effect of the equivalent of a commute between “home” and “work”, or “suburban” and “urban”, to obtain an effective $R_0$, which is found to be equal to the arithmetic mean weighted by the exposure time.  

We subsequently considered the effect of \rededit{mobility, first by considering diffusion in 1-D geometries. We showed the emergence of infection waves in rectilinear and also in radial geometries, which asymptotically travel with constant velocity, the dimensional value of which scales with the square root of diffusivity, and increases with $R_0$. The behavior of the infection curves at a fixed time is similar to that for an SIR system, which they approach at  relatively large values of $R_0$. We then examined how distributed densities in different geometries affect the propagation of infection waves. We find that for all practical purposes, the contagion propagation can be modeled equivalently by wave propagation, the speed of which only depends on the value of $R_0$ in that region. Finally, we explored aspects of advection in the case of a stochastic velocity distribution. Advection leads to an effective macrodispersion, which dominates over diffusion, and leads to enhanced "mixing", hence effectively to enhance spreading of the infection. More work is required to fully explore these effects.} Spatial effects via density, diffusion or advection considerations are important and must be considered in the description of contagion and its epidemics.  

\rededit{The fundamental purpose of this work was to provide a robust framework that allows the modeling of the spreading of human-to-human infections by accounting for mobility (diffusion and advection) and chemical reaction-like attributes. A number of new insights have been uncovered, from how to represent the kinetic parameters and how to include density dependence, to the role of diffusion and advection in the spatial propagation of infection. This has captured essential, although not all relevant variables. For example, limiting the various sub-populations to three ignores important demographic attributes as well as the possibility of correlations in infections, e.g. through family or work relationships. Reflecting human behavior, the important parameter $R_0$ can vary in space and time. Such variations were considered in many of our applications. Solving the inverse problem of matching real data has not been attempted in this work, other than in a qualitative way. This task is to be considered next, by possibly extending the approach presented to additional populations. Regardless, we believe that the approach presented here provides a fundamentally sound framework, based on extending fundamental physico-chemical principles to model collective human behavior, and should be useful to the further understanding of the spreading of infections.}             
\newpage

\bibliographystyle{unsrt}  
\bibliography{references}  


\end{document}